\documentclass[prb,twocolumn,floats,showpacs]{revtex4}
\usepackage{amsmath}
\usepackage{bm}

\usepackage{graphicx}
\usepackage{subfigure}
\usepackage{epsfig}
\usepackage{psfrag}
\usepackage{amssymb}

\newcommand{\nt}{carbon nanotube}
\newcommand{\rlv}{reciprocal lattice vector}
\newcommand{\me}{\mathrm{e}}
\newcommand{\mi}{\mathrm{i}}
\newcommand{\modulus}[1]{\left| #1 \right|}
\newcommand{\ul}{\mathbf}

\newcommand{\expect}[3]{\left< #1 \right| #2 \left| #3 \right>}

\newcommand{\ket}[1]{\left| #1 \right>}

\newcommand{\be}{\begin{equation}}
\newcommand{\ee}{\end{equation}}

\begin{document}

\title{Effects of Disorder and Momentum Relaxation on the Intertube
Transport of Incommensurate Carbon Nanotube Ropes and Multiwall
Nanotubes}

\author{M. A. Tunney and N.R. Cooper}
\affiliation{T.C.M. Group, Department of Physics,
Cavendish Laboratory,
J. J. Thomson Avenue, Cambridge CB3 0HE, United Kingdom.}



\date{1 May 2006}

\begin{abstract}

We study theoretically the electrical transport between aligned carbon
nanotubes in nanotube ropes, and between shells in multiwall carbon
nanotubes. We focus on transport between two metallic nanotubes (or
shells) of different chiralities with mismatched Fermi momenta and
incommensurate periodicities. We perform numerical calculations of the
transport properties of such systems within a tight-binding
formalism. For clean (disorder-free) nanotubes the intertube transport
is strongly suppressed as a result of momentum conservation.  For
clean nanotubes, the intertube transport is typically dominated by the
loss of momentum conservation at the contacts.  We discuss in detail
the effects of disorder, which also breaks momentum conservation, and
calculate the effects of localised scatterers of various types.  We
show that physically relevant disorder potentials lead to very
dramatic enhancements of the intertube conductance.  We show that
recent experimental measurements of the intershell transport in
multiwall nanotubes are consistent with our theoretical results for a
model of short-ranged correlated disorder.

\end{abstract}

\maketitle

\section{Introduction}

The electronic properties of carbon nanotubes are the source of much
very interesting physics. Not only are \nt s highly one-dimensional
conductors, but the dependence of the nanotube bandstructure on the
the nanotube's ``chirality'' leads to a range of different
bandstructures: both metallic and semiconducting behaviours arise and,
even within each of these, the possible bandstructures vary
considerably.\cite{saito} The nature of the bandstructure has
important effects on the electronic transport properties, not least
the large difference in the effects of disorder on metallic nanotubes
as compared to (doped) semiconducting
nanotubes.\cite{andohelicity1,andohelicity2,mceuen}

Theory shows that the bandstructure also plays key roles in the
intertube transport between (the outer shells of) aligned nanotubes in
a rope,\cite{maarouf} and in the transport between the shells of a
multiwall nanotube (MWNT).\cite{roche,louie,ahn,uryu1,triozon}
[Henceforth, we shall discuss these two cases together by referring to
the transport as inter{\it tube} transport, with the two ``tubes'' to
be understood as single-wall nanotubes (SWNTs) representing either the
outer shells of two neighbouring nanotubes in a nanotube rope, or two
neighbouring shells in a MWNT.]  In experimental systems of this type,
the two tubes typically have differing chiralities and therefore
differing bandstructures.  Even when the two tubes are both metallic,
the typical case poses a complicated situation: the Fermi momenta do
not coincide, and their translational periods are
incommensurate. Consequently, as a result of momentum conservation,
one expects a strong suppression of intertube tunnelling\cite{maarouf}
as compared to the simplest case of two identical tubes (with matched
Fermi momenta and commensurate periodicity).  Indeed, theory has led
to the counter-intuitive prediction that increasing the length over
which the two shells of a MWNT are coupled will {\it reduce} the
intershell conductance.\cite{louie}

A knowledge of the intertube conductance in nanotube ropes and
intershell conductance in MWNTs is of fundamental importance for
understanding the transport mechanisms in these systems. However,
direct experimental studies of these quantities cannot easily be
achieved. Very elegant recent experimental work has succeeded in
probing the intertube/shell transport. Experimental studies on
intentionally disordered nanotube ropes\cite{stahl} have been been
used to estimate the intertube tunnelling conductance.  Non-local
resistance measurements between contacts on the outer shells of MWNTs
have been used to determine the intershell conductivity.\cite{bourlon}
(The development of telescopically extended MWNTs\cite{cumings} opens
the way to future studies of intershell conductance. However, at
present the transport properties appear not to be dominated by the
intershell conductance.\cite{cumings})

As emphasised in Ref.~\onlinecite{bourlon}, the experimental
measurements of intershell transport in MWNTs cannot be accounted for
within existing theories. The experimental systems have some degree of
disorder, and existing theories of incommensurate tubes do not take disorder into
account.\cite{roche,louie,ahn,uryu1,triozon} Disorder is likely to have a very dramatic effect on the
intertube/shell transport: disorder breaks translational symmetry and
therefore relaxes the momentum conservation which is responsible for
the suppression\cite{maarouf,louie} of intertube transport.

In this paper, we report calculations of the effects of disorder on
the intertube conductance in carbon nanotube ropes and MWNTs.  In the
rope geometry, we consider two nearest-neighbour tubes and study the
transport between the outer shells of these tubes. In the multiwall
geometry (in which the shells are coaxial) we study the transport
between two neighbouring shells of the tube [thus we effectively
consider a double wall nanotube (DWNT)]. We focus on two metallic
shells of armchair and zigzag type, which provides an example of the
typical case involving mismatched Fermi momenta and incommensurate
periodicities.  We show that defects lead to very dramatic {\it
enhancements} of the intertube conductance. We quantify the effects of
defects of different types, and discuss the results in connection with
the experimental results of Refs.~\onlinecite{stahl} and
\onlinecite{bourlon}. Our results show that disorder has a large
effect for both the intertube transport in disordered ropes measured
in Ref.~\onlinecite{stahl}, and for the intershell transport in MWNTs
measured in Ref.~\onlinecite{bourlon}.  By comparing our theory to the
experimentally determined intershell conductance of
MWNTs\cite{bourlon} we put constraints on the nature of the disorder
in these samples, and find parameter regimes of disorder for which our
theoretical results are consistent with the experimental
measurements.\cite{bourlon}

The paper is organised as follows. In \S\ref{sec:model} we describe
the model that we use, and discuss the importance of translational and
rotational symmetries in the intertube coupling in aligned \nt s. We
introduce a classification of the coupling strength between two shells
of a MWNT based on the rotational symmetry.  In \S\ref{sec:gf} we
describe the Green's function approach used in our numerical
studies. In \S\ref{sec:clean} we apply this to the case of clean
(disorder-free) nanotubes. We demonstrate the importance of momentum
conservation for clean nanotubes.  We show that momentum-relaxation at
the contacts is the dominant effect controlling the intertube
transport in clean nanotubes.  The importance of our classification of
the rotational symmetry in MWNTs is demonstrated through calculations
of transport between nanotube shells in a telescopic junction
geometry.
In \S\ref{sec:defects} we turn to study the effects of disorder on the
intertube conductance. We calculate the effects of localised defects
of different types, and compare our results to the experimental
measurements of Refs.~\onlinecite{stahl} and \onlinecite{bourlon}.
Finally in \S\ref{sec:conclude} we summarise the main results of the
paper.

While preparing this work for publication, we learned of related
work\cite{uryu} where the importance of momentum relaxation at the
contacts of clean DWNTs was also recognised. Since our discussion of
this aspect in \S\ref{sec:clean} is somewhat different from
Ref.~\onlinecite{uryu}, and since an understanding of the clean case
is required in order to introduce our calculations of the effects of
defects, we retain this discussion.  Our results in this section are
consistent with those of Ref.~\onlinecite{uryu}.

\section{Nanotube Bandstructures, Translational and Rotational Symmetries}

\label{sec:model}

\subsection{Theoretical Model}

We describe the bandstructure of the nanotubes using a one-band tight
binding Hamiltonian,
\begin{equation}
\label{eqn:tightbindingham}
\mathcal{H}_0 = -t \sum_{\langle i,j\rangle ,\sigma}\left(
c^{\dag}_{i\sigma}c_{j\sigma}+c^{\dag}_{j\sigma}c_{i\sigma}\right)
\end{equation}
where $c_{i\sigma}$ and $c_{i\sigma}^{\dagger}$
are the annihilation and creation operators for an electron with spin
polarisation $\sigma = \pm 1/2$ on the atomic site $i$. The atoms lie
on a (rolled up) honeycomb lattice, with lattice constant
$a=2.46\mbox{\AA}$ and hopping occurring only between nearest
neighbour sites $\langle i,j\rangle$.  For quantitative estimates, we
use $t=2.77\mbox{eV}$ consistent with the value fitted to the results
of tunnelling experiments.\cite{wilder}


We consider intertube conductance between two \nt s due to the
addition of a tunnelling Hamiltonian
\begin{equation}
\mathcal{H}_{T}=-\sum_{j,i,\sigma}
t_{ij}\left(c^{\dagger}_{i\sigma}c_{j\sigma}+
c^{\dagger}_{j\sigma}c_{i\sigma}\right)
\label{eq:tunnel}
\end{equation}
where $i$ and $j$ now label atomic sites on two different tubes.
Following Ref.~\onlinecite{maarouf}, we model the intertube hopping
(between atoms at positions $\bm{r}_i$ and $\bm{r}_j$) by
\begin{equation} t_{ij}= t_{\perp}\me^{-|\bm{r}_i-\bm{r}_j|/{\delta}}
\label{eqn:maarparams}
\end{equation}
with $t_{\perp} =492\mbox{eV}$ and $\delta = 0.5\mbox{\AA}$.  We have
also performed calculations (not reported here) using a more
sophisticated model\cite{ando} which accounts for the angular
dependence of the hopping through $\sigma$ and $\pi$ bonds.  We find
similar results using either parameterisation in all cases we have
studied (with realistic geometries).  For the purposes of this paper,
we therefore restrict attention to the results of the simpler model
(\ref{eqn:maarparams}).

In all the calculations we report, we treat the electrons as
non-interacting.  The results are thus valid at high temperatures, or
for systems with many occupied modes, as in MWNTs, where the effects
of interactions are suppressed. We discuss briefly the influence of
interactions in \S\ref{sec:conclude}.

\subsection{Bandstructures of Armchair and Zigzag Nanotubes}

\label{subsec:aczz}

It is well-known that the electronic properties of carbon nanotubes
depend on the chiral vector $(n_1,n_2)$, which sets the quantised
subbands and thus the bandstructure.\cite{saito} In this work we focus
on metallic armchair and zigzag nanotubes.

Armchair tubes are described by $(n_a,n_a)$ chiral vectors.  The
wavevector around the tube circumference takes the values $k_\perp =
(2\pi/\sqrt{3}a) (q/n_a)$ with $q=1,2,\ldots 2n_a$.  For each $q$
there are two bands, labelled by $\pm$, which have the energy
dispersions
 \begin{equation}
\epsilon_{q}^{\rm ac}(k)=\pm t
\left[1+4\cos{\left(\frac{q\pi}{n_a}\right)}\cos{\left(\frac{ka}{2}\right)}+4\cos^{2}{\left(\frac{ka}{2}\right)}\right]^{\frac{1}{2}}
\label{eqn:acdisp}
\end{equation}
where $k$ is the wavevector along the tube, with $-\pi < ka <
\pi$.  The $\pm$ bands with $q=n$ are metallic, crossing $E=0$ at $k=
\pm 2\pi/(3a)$.

Zigzag tubes are described by $(n_z,0)$ chiral vectors. The 
wavevector around the tube circumference takes values $k_\perp = (2\pi/a)
(q/n_z)$, with $q=1,...,2n_z$, and the subbands have dispersion relations
\begin{equation}
\epsilon_{q}^{\rm zz} (k)=\pm t
\left[1+4\cos{\left(\frac{\sqrt{3}ka}{2}\right)}\cos{\left(\frac{q\pi}{n_z}\right)}+4\cos^{2}{\left(\frac{q\pi}{n_z}\right)}\right]^{\frac{1}{2}}
\label{eqn:zzdisp}
\end{equation}
with $-{\pi}/{\sqrt{3}}<ka<{\pi}/{\sqrt{3}}$.  If $n_z$ is an
integer multiple of $3$, the zigzag tube is metallic.  The
metallic subbands have $q={2n_z}/{3}, {4n_z}/{3}$ (the $\pm$ bands
correspond to right- and left-moving modes of these two subbands).
The two metallic subbands both cross $E=0$ at $k=0$.

\begin{figure}
\centering
\includegraphics[scale=0.3]{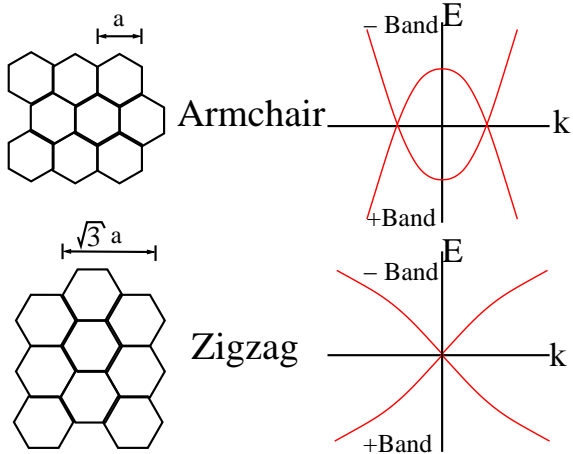}
\caption{(Color online) Schematic diagrams of the lattice structures (the tube axes
run horizontally) and bandstructures of the metallic bands for
armchair and zigzag nanotubes.  The repeat distances along the tube
axes differ by a factor of $\sqrt{3}$, so the tubes are
incommensurate.  Intertube tunnelling will be suppressed for Fermi
energies close to $E=0$, owing to the lack of energy-momentum
conserving processes.}
\label{fig:incombands}
\end{figure}

The lattice structures and metallic bands of armchair and zigzag
nanotubes are illustrated in Fig.~\ref{fig:incombands}.

\subsection{Intertube Conductance: Momentum Conservation and Incommensurability}

\label{subsec:theoryincommens}

We explore the effects of momentum (wavevector) conservation in the
intertube conductance between two aligned nanotubes.  First, in order
to explain the importance of momentum conservation, we consider the
tubes to be sufficiently weakly coupled that this coupling can be
described by perturbation theory.  This is valid provided the
intertube conductance is small compared to $e^2/h$. (Our numerical
results will not be restricted to this perturbative limit.)  Within
perturbation theory, for transport between a Fermi point $k_{\rm A}$
on one tube and a Fermi point $k_{\rm B}$ on the other there is a
contribution to the intertube conductance of
\begin{equation}
G_{k_{\rm A},k_{\rm B}}=
  \frac{2e^{2}}{\hbar^{3}v_{\rm A}v_{\rm B}}\frac{L_N^{2}}{2\pi}
\modulus{t_{k_{\rm A} k_{\rm B}}}^{2}
  \label{eqn:goldenrulecond1}
\end{equation}
where
\begin{equation}
\label{eq:tunme}
t_{k_{\rm A}k_{\rm B}}   \equiv  \langle k_{\rm B}|{{\mathcal{H}_T}}|{k_{\rm A}}\rangle
\end{equation}
is the tunnelling matrix element, $v_{\rm A}$ and $v_{\rm B}$ are the
velocities at the two Fermi points, and $L_N$ is the length over which
the states $\ket{k}$ are normalised.  The total conductance is
obtained by the sum of (\ref{eqn:goldenrulecond1}) over all initial
and final Fermi points ($k_{\rm A},k_{\rm B}$) associated with the
different occupied subbands [we have suppressed band indices in
Eqns.~\ref{eqn:goldenrulecond1} and~\ref{eq:tunme}].

Momentum conservation arises from the symmetry of the tunnelling
Hamiltonian (\ref{eqn:maarparams}) under translation along the
tube.\cite{footnote1}
If the Fermi momenta of the two tubes are the same, ($k_{\rm A} =
k_{\rm B}$), then electrons at the Fermi energy can tunnel between the
tubes conserving both energy and momentum: one expects a large
intertube conductance. (For two tubes of the same type, the Fermi
momenta coincide at all energies so intertube coupling is always
strong.)  If the Fermi momenta differ ($k_{\rm A}\neq k_{\rm B}$),
tunnelling will be suppressed.
In general, tubes of different chiralities have different dispersion
relations and therefore differing Fermi momenta.  For example, as
shown in Fig.~\ref{fig:incombands}, the Fermi momenta of the armchair
and zigzag tubes differ in a range of Fermi energies around $E=0$.
(At $E_F=0$ the Fermi momenta are $k_F^{\rm ac} = \pm 2\pi/(3a)$ and
$k^{\rm zz}_F = 0$ for armchair and zigzag tubes respectively.)

Although one might conclude that if the Fermi momenta differ,
intertube tunnelling will vanish identically, this is not always true.
An important distinction arises between the cases in which the tubes
have {\it commensurate} or {\it incommensurate} lattice structures.
Denoting the repeat distances in the two tubes by $a_{\rm A}$ and
$a_{\rm B}$, we say that the \nt s are commensurate if $a_{\rm
A}/a_{\rm B}$ is rational fraction, and that they are incommensurate
if this is an irrational fraction.

If the \nt s are {\it commensurate}, then the system has translational
symmetry along its length. The minimum distance along the tubes under
which the system is translationally invariant sets the size of the
first Brillouin zone.  If the Fermi momenta of the two tubes do not
coincide within this first Brillouin zone, then there can be no
tunnelling between the tubes that conserves energy and (crystal)
momentum.  The intertube tunnelling will vanish.

If the \nt s are {\it incommensurate}, then there is no finite
distance under which the system is translationally invariant: the size
of the first Brillouin zone is vanishingly small, and the eigenstates
cannot be characterised by a conserved crystal momentum.
Nevertheless, for weak coupling between the tubes, the Bloch states of
the individual tubes still provide a useful starting point.  Imagine
that, in the first Brillouin zones of the individual tubes, the Fermi
momenta of the two tubes do not coincide.  Because the \rlv s are
incommensurate, by adding sufficient \rlv s to each of the tubes one
will be able to bring the Fermi momenta arbitrarily close. (That is,
in an extended zone scheme the Fermi momenta can become arbitrarily
close.)
However, as shown by Maarouf {\it et al.}~\cite{maarouf}, the
larger the number of \rlv s that must be added to achieve momentum
matching, the weaker is the tunnelling between the two tubes.  
Ref.~\onlinecite{maarouf} provided an approximate calculation of the matrix
element of the intertube coupling (\ref{eq:tunnel}) between momentum
eigenstates on two aligned nanotubes placed side-by-side (as in a
carbon nanotube rope).  Following this approach,\cite{maarouf} which
is valid in the limit that tunnelling is dominated by the points of
closest approach, we find that the matrix element is
\begin{widetext}
\begin{eqnarray}
t_{\ul{k}^{\rm A},\ul{k}^{\rm B}}
 & = &
t_{\perp}\delta_{k^{\rm A}_{*},k^{\rm B}_{*}}\sqrt{\frac{2\pi}{b+2R}}
\frac{b\delta^{\frac{3}{2}}}{A_{\rm cell}}\me^{-{b}/{\delta}}
\me^{-{b\delta K^{2}}/{2}}
\me^{-{\left(k^{\rm A}_{\perp}-k^{\rm B}_{\perp}\right)^{2}R\delta}/{4}}
\me^{-{Rb\delta\left(k^{\rm A}_{\perp}+k^{\rm B}_{\perp}\right)
^{2}}/{4\left(b+2R\right)}}
\label{eqn:maarouftkk}
\end{eqnarray}
\end{widetext}
where $A_{\rm cell}$ is the area of a unit cell, $k^{\rm A,B}_\perp$
are the (subband) wavevectors around the tube circumference, $k^{\rm
A}_{*}\equiv k^{\rm A}+\alpha G^{\rm A}$ and $k^{\rm B}_{*}\equiv
k^{\rm B}+\beta G^{\rm B}$ are the wavevectors along the tubes shifted
by integer numbers ($\alpha,\beta$) of \rlv s along the tube ($G^{\rm
A}, G^{\rm B}$), $K^{2}\equiv \modulus{k^{\rm
A}_{*}}^{2}+\modulus{k^{\rm B}_{*}}^{2}$, and $b$ is the distance
between tubes at the point of closest approach.  For larger numbers of
\rlv s, $K^2$ increases and the tunnelling matrix element is
exponentially suppressed.  One expects the tunnelling to be dominated
by processes involving the smallest number of additional \rlv
s.\cite{maarouf}

Two tubes of different chiralities typically have Fermi momenta that
do not coincide, and typically are incommensurate.  The armchair and
zigzag tubes are an example of this typical case. Here, the smallest
reciprocal lattice vectors are $G^{\rm ac} = 2\pi/a$, $G^{\rm zz} =
2\pi/\sqrt{3}a$. The condition for energy-momentum conservation is
\begin{equation}
\epsilon_{q_a}^{\rm ac}\left(k_F^{\rm ac}+\alpha
G^{\rm ac}\right)=\epsilon_{q_z}^{\rm zz}\left(k_F^{\rm zz}+\beta G^{\rm zz}\right) = E_F
\label{eqn:peakcondition}
\end{equation}
where $k_F^{\rm ac/zz}$ are the Fermi momenta along the respective
tube axes, and $\alpha$ and $\beta$ are integers. Focusing on the
dominant tunnelling process involving just one reciprocal lattice
vector of each tube, we look for a solutions with $\alpha=\pm 1$ and
$\beta=\pm 1$. Using the dispersion relations
(\ref{eqn:acdisp},\ref{eqn:zzdisp}) we find $ka=\pm 3.914$ at energies
$E_F=\pm E_0$ where $E_0 = 0.247 t$.  We shall show the importance of
this energy in the numerical results in \S\ref{subsec:ropegeom}.

\subsection{Multiwall Geometry: Rotational Symmetry}

\label{sec:mwsym}

In the case of MWNTs, intertube conductance is strongly affected not
only by momentum conservation along the tube, but also by angular
momentum conservation around the axis of the tube.  This leads to
additional selection rules in the tunnelling matrix elements which
depend sensitively on the chiralities of the two tubes.\cite{louie,kimchang}

We discuss the implications of this symmetry for the tunnelling matrix
element (\ref{eq:tunme}) between armchair and zigzag shells.  The
subband wavefunctions vary as $\me^{\mi k_\perp y}$, where $y$ is the
coordinate around the circumference of the tube.  For an $(n_a,n_a)$
armchair tube, $k_{\perp}={2\pi}/({\sqrt{3}a})({q_{a}}/{n_{a}})$, and
the lattice is invariant under a translation by $\sqrt{3} a$ in $y$. Using
$\theta_i^a = y_i/(\sqrt{3} an_a) = i/n_a$ to label the angular
position of the unit cell $i$ ($i=1,\ldots n_a$), then the subband
wavefunction is simply $\me^{\mi q_a \theta_i^a}$. Similarly for the
zigzag tube, $\theta^z_j = y_j/(a n_z) = j/n_z$, and the subband
wavefunction is $\me^{\mi q_z \theta_j^z}$ for the unit cell labelled by
$j=1,\ldots n_z$.  Therefore, the tunnelling matrix element contains an
angular contribution of the form
\begin{align}
\expect{\Psi_{z}}{\mathcal{H}_{T}}{\Psi_{a}} &\propto
\sum_{i,j}\tilde{t}(\theta^a_{i}-\theta^z_{j}) \me^{\mi q_{a} \theta^a_{i}}
\me^{-\mi q_{z} \theta^z_{j}} \nonumber \\
&=\sum_{i,j}\sum_{n}\tilde{t}_{n}\me^{-\mi n (\theta^a_{i}-\theta^z_{j})}
 \me^{\mi q_{a} \theta^a_{i}}
\me^{-\mi q_{z} \theta^z_{j}} \nonumber
\end{align}
where $\tilde{t}(\theta^a_i-\theta^z_j)$ is the function  obtained by
expressing (\ref{eqn:maarparams}) in terms of the angular separation
between atoms in the two shells, which in the last line has been
expressed as a Fourier series due to its periodicity. Performing the
sums over $i$ and $j$, one finds that the contribution of the harmonic
$\tilde{t}_n$ is nonzero only if
\begin{eqnarray}
\label{eqn:modconditiona}
\text{Condition A:}\quad \mbox{mod}(n-q_a,n_a) &=&0\\
\text{Condition B:}\quad \mbox{mod}(n-q_z,n_z) & = & 0
\label{eqn:modconditionb}
\end{eqnarray}
If both of these conditions are met, then modes $q_{a}$ and $q_{z}$
are connected by a non-zero tunnelling matrix element. Restricting
attention, now, to the metallic bands in both tubes -- that is,
$q_a=n_a$  and $q_z = 2n_z/3$ or $q_z = 4n_z/3$ -- the conditions
(\ref{eqn:modconditiona}), (\ref{eqn:modconditionb}) may be rewritten
\begin{eqnarray*}
\text{A:}\quad n&=&\alpha n_{a} \\
\text{B:}\quad n&=&n_{z}\left( \frac{1}{3}+\beta \right)
\quad \mbox{or}\quad n=n_{z}\left( \frac{2}{3}+\beta \right)
\end{eqnarray*}

We classify the nature of the rotational symmetry of the tubes in
three cases. Which case applies depends on the ratio ${n_{a}}/{n_{z}}
= {p}/{q}$ where $p$ and $q$ are natural numbers with no common
factor.

 (i){\bf ``Zero''}: No value of $n$ that satisfies condition B
satisfies A.  This is the case if $q$ is not divisible by $3$.  The
tunnelling matrix elements between the metallic bands of the armchair
and zigzag tubes vanish identically.

(ii) {\bf ``Strong''}: All values of $n$ that satisfy condition B,
satisfy A.  This is the case if $q$ is divisible by $3$ and if $p=1$.
The metallic bands of the armchair and zigzag tubes couple strongly,
containing contributions from  all possible angular harmonics of the
tunnelling operator.

(iii) {\bf ``Intermediate''}: Some values of $n$ that satisfy
condition B satisfy A. This is the case if $q$ is divisible by $3$ and
if $p\neq 1$.  The metallic bands of the armchair and zigzag tubes are
coupled, but contain contributions from only a subset of the angular
harmonics of the tunnelling operator.

That rotational symmetry can lead to vanishing matrix elements (the
``zero'' case) is well-known.\cite{louie,kimchang} However, as we show
below in our numerical results, there can be very substantial
quantitative differences between cases we discriminate as
``intermediate'' or ``strong'' coupling.

\section{Conductance using Green's Functions}

\label{sec:gf}

\begin{figure}
\centering
\includegraphics[scale=0.3]{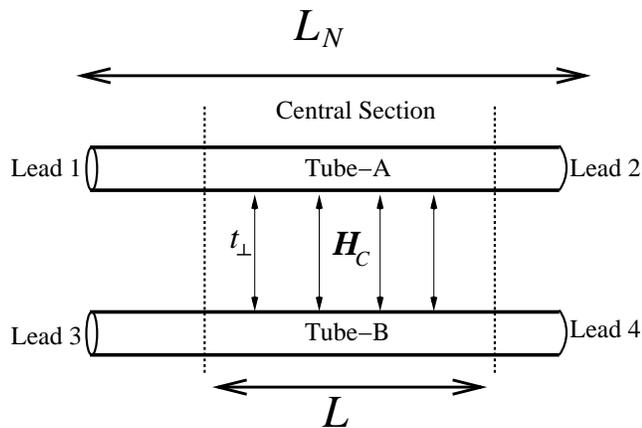}
\caption{Geometry of the system that we study, illustrating the
labelling of the leads and the length $L$ of the central section where
intertube tunnelling can occur. The length $L_N$, which includes the
leads, is used for normalisation and is assumed large.}
\label{fig:sharpwires}
\end{figure}

In our numerical studies, we calculate the conductance of the double
nanotube system within the Landauer-B{\" u}ttiker framework, with the
scattering matrix calculated numerically using a Green's function
approach.\cite{datta} The geometry of the system studied is shown in
Fig.~\ref{fig:sharpwires}.
It consists of two parallel nanotubes, each of which we shall consider
to be either armchair or zigzag type.  The geometry allows a
description of (the outer shells of) two neighbouring tubes in a rope
geometry (if the tube axes do not coincide), or two neighbouring
shells in a MWNT (if the axes do coincide).

We calculate the conductance using a scattering approach.  For this
purpose, we consider leads to be attached to the ends of each
nanotube.  The leads are ideal, consisting of a semi-infinite section
of clean nanotube, and there is no intertube coupling between the
leads.  The intertube coupling is confined to the central section of
length $L$. In this region the system has Hamiltonian $\mathcal{H}_C$,
which describes atomic sites that experience intra- and inter-tube
hopping and where there can be diagonal or off diagonal disorder. The
nature of the intertube scattering in this region determines if we are
considering the rope or multiwall geometry.

Following the approach described in Ref.~\onlinecite{datta}, the
overall Green's function at energy $E$ can be obtained via a matrix
inversion
\begin{equation} \label{eqn:gcmatr}
\mathcal{G}_{C}=\left[E \mathcal{I}-\mathcal{H}_{C}-\Sigma\right]^{-1}
\end{equation}
where the matrix has dimension $C\times C$, with $C$ the number of
sites in the central region. [The typical system sizes we study have on
the order of $C=9000$ atoms. Matrix inversion is
done using standard LaPack code.]  The effects of the
leads are taken into account via a self-energy term
$\Sigma=\Sigma_{p}\mathcal{\tau}^{\dag}_{p}g_{p}\mathcal{\tau}_{p}$
which can be viewed as an effective Hamiltonian describing the coupling
between the conductor and the four leads (labelled by $p=1,\ldots 4$).
In this expression, $g_{p}$ is the end Green's function of lead $p$
and $\tau_p$ is a matrix describing the coupling of this lead to the
central region. Although the leads are semi-infinite, the end Green's
function has finite dimension, set by the number of sites on the end of the
lead.

We restrict attention to the case in which the energy is sufficiently
close to $E=0$ that there are only two conducting modes in each of the
metallic nanotube leads. This allows a description of transport
properties when the energy $E$ is less than the gap to the next
subband.
In this case, it is sufficient to calculate the Green's function for
the leads including only these metallic modes.  We have determined the
end Green's function analytically for the metallic modes of the
armchair and zigzag tubes (the calculation is described in
Appendix~\ref{app:ntgreenfns}), allowing an analytic determination of
the self-energy entering Eqn.~\ref{eqn:gcmatr}.  Note that in the
``central'' region the system is solved fully (numerically), including
all subbands.

For the range of energy that we study, the transmission matrix is an
$8\times 8$ matrix, $T_{\mu,p; \mu',p'}$, describing the transmission
from two incoming modes (labelled by $\mu$) in each of the four leads
($p=1,2,3,4$) to the $2\times 4$ outgoing modes (with mode $\mu'$ and
lead $p'$) The nature of the mode depends on whether the lead is of
armchair or zigzag type: for an $(n_a,n_a)$ armchair tube, $\mu$
labels the $\pm$ bands with $q_a=n_a$; for a $(0,n_z)$ zigzag tube,
$\mu$ labels $q_z=2n_z/3, 4n_z/3$.  We obtain the scattering matrix
for currents, following the approach described in
Ref.~\onlinecite{datta} for a continuum system applied to this
discrete case. As in the case of the continuum, it is important to
convert the matrix for scattering amplitudes into the matrix for
scattering fluxes using the velocities of the initial and final modes.
From the scattering amplitude for the flux, one obtains the
transmission probabilities, the final results for which are given in
Appendix\ref{app:transmissioncoefficients}.

Finally, using the
Landauer-B{\" u}ttiker approach, the conductance is obtained from the
sum of the transmission probabilities.  For example, the intertube
conductance between tube-A ($p=1,2$) and tube-B ($p=3,4$) as measured
in an experiment with contacts attached to all four leads, involves
the total transmission coefficient between all modes in the tube-A and
all modes in tube-B \be G = \frac{2e^2}{h}
\sum_{\mu;p=1,2}\sum_{\mu';p'=3,4} T_{\mu,p; \mu', p'}
\label{eq:ginter}
\ee
The prefactor of 2 arises from spin degeneracy. The maximum
intertube conductance in this case is $8 e^2/h$.

\subsection{Application: Conductance of a SWNT with a Vacancy}
\label{sec:singletube}

As a first application of our numerical method, we study the
two-terminal conductance of {\it single} SWNTs (both armchair and
zigzag) in the presence of a vacancy. This is achieved simply by
turning off intertube tunnelling, $t_\perp=0$, and calculating the
scattering between the two ends of the individual tubes. Thus, the
two-terminal conductance of tube-A is
$$G_{\rm intra} = \frac{2e^2}{h} \sum_{\mu,\mu'}
T_{\mu,1; \mu', 2}$$
which has maximal value ${4e^{2}}/{h}$ for a clean metallic nanotube (two
propagating modes and twofold spin degeneracy).

Related calculations of the effects of vacancies on armchair SWNTs
have been reported previously.\cite{hansson2,choi,chico,igami} As in
those studies we do not include relaxation of the lattice around the
vacancy, but leave the honeycomb lattice undistorted.
Following
Ref.~\onlinecite{igami} we model the vacancy by imposing a huge
potential on the vacancy site ($10^8\mbox{eV}$), which was
shown\cite{igami} to be equivalent to explicit disconnection of the
bonds to this site.

\begin{figure}
\centering
\includegraphics[scale=0.3]{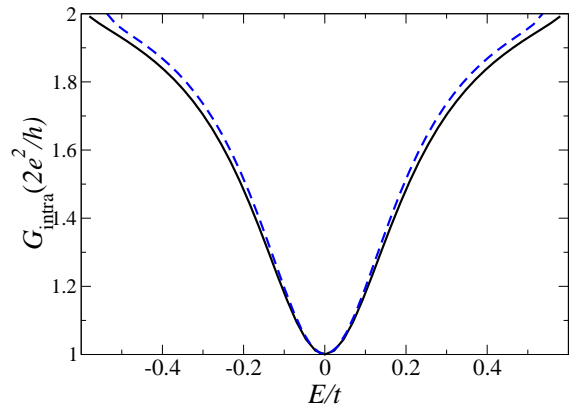}
\caption{(Color online) Two-terminal conductance of a $(5,5)$ armchair nanotube with a single
  vacancy (solid line) and a (9,0) nanotube with a single vacancy (dashed line). The tube length ($L=31a$) used in these calculations is sufficiently long that the results are characteristic of the bulk behaviour.}
\label{fig:singtubevac}
\end{figure}

In Fig.~\ref{fig:singtubevac} we present results for the conductance
of single metallic armchair and zigzag \nt s in the presence of a
single vacancy.
Our calculations reproduce the results previously found by Igami {\it
et al.}\cite{igami} for armchair tubes.  In addition, we present
results for a metallic zigzag nanotube.
The results in both cases (armchair and zigzag) are very similar: the
conductance drops from close to its ideal value of ${4e^{2}}/{h}$ at
the edge of the second conduction band, to ${2e^{2}}/{h}$ at $E=0$.\cite{footnote2}

One can understand that the minimum value of the conductance at $E=0$
is $G_{\rm intra} = 2e^2/h$ by noting that, in both armchair and
zigzag cases, a linear combination of the two degenerate right-moving
(or left-moving) states of an ideal (defect free) nanotube can be
formed that has zero amplitude on a subset of the lattice sites.
Choosing the linear combination such that the amplitude is zero on the
site containing the vacancy, this mode of the ideal tube is unaffected
by the presence of a vacancy and therefore contributes
$2e^2/h$ to the two-terminal conductance. That our numerical results
show that the conductance falls to precisely this minimum value at
$E=0$  indicates that the other (orthogonal) mode is
completely reflected by the vacancy.
For an armchair tube, the wavefunction of the transmitted mode is
periodic along the length of the tube every three lattice constants
[it is a superposition of waves with $k=+2\pi/(3a)$ and
$k=-2\pi/(3a)$].  Therefore, one expects that it should be possible to
introduce several vacancies, at positions chosen judiciously to lie at
the set of zeroes of this transmitted mode, without having any effect
on its transmission.  We have checked this expectation numerically by
introducing several vacancies in the \nt~on sites respecting the
periodicity of the wavefunction: at $E=0$ the conductance remains
${2e^{2}}/{h}$, indicating that the transmitted wave indeed does not
feel the potentials on these sites.

\section{Intertube conductance between two clean SWNTs}

\label{sec:clean}

We now turn to consider the intertube conductance between aligned
nanotubes in both rope and multiwall geometries.  From the discussions
of \S\ref{subsec:theoryincommens} we expect that momentum conservation
and, in the case of MWNTs, the rotational symmetry will play important
roles in determining the magnitude of the intertube conductance. Here,
we investigate these effects, focusing on the case of clean tubes in
the absence of defects. A discussion of the effects of defects will be
given in \S\ref{sec:defects}.  We begin our discussion by
considering the effects of the loss of momentum conservation at the
contacts.

\subsection{Contact Effects}

\label{subsec:contacts}

The presence of current contacts at the two ends of a carbon nanotube
sets a finite total length of the system, $L$.  In
Ref.~\onlinecite{louie} it was recognised that as a consequence of
this finite length, $L$, momentum conservation will be relaxed on a
wavevector scale of order $1/L$. This led to the
prediction\cite{louie} of a decrease in the intershell conductance of
a MWNT with increasing length $L$, as momentum conservation becomes
more accurately satisfied.

Another effect is that in the vicinity of the contacts themselves the
translational symmetry is broken.
We denote
the (minimum) intrinsic lengthscale of the contact $L_{c}$: this could
be the size of the contact, or the lengthscale over which the
electronic nanotube density changes between a region under the contact
and a region away from the contact. The variation of the nanotube
properties on $L_c$ will promote a loss of momentum conservation up to
a wavevector scale of about $1/L_c$. Since  $1/L_c\gg 1/L$,
this change in momentum will typically be larger than that arising
from the finite length of the tube.
We find that momentum relaxation at the contacts typically dominates
the transport properties between clean nanotubes with mismatched Fermi
momenta.
The contact-induced contribution to the intertube conductance was not
considered in Ref.~\onlinecite{louie}.

Since we are interested in the general features of momentum (non-)
conservation at the contacts, we introduce a simple model that
incorporates a contact lengthscale $L_c$ as well as the total length
$L$.  Specifically, we replace (\ref{eqn:maarparams}) by
\begin{widetext}
\begin{equation}
t_{ij}=t_{\perp}\me^{-{|\bm{r}_i-\bm{r}_j|}/{\delta}}\times
f\left(\frac{x_{i}-L}{L_{c}}\right)\times
f\left(-\frac{x_{i}}{L_{c}}\right)\times
f\left(\frac{x_{j}-L}{L_{c}}\right)\times
f\left(-\frac{x_{j}}{L_{c}}\right)
\label{eq:trirj}
\end{equation}
\end{widetext}
where $f(x) \equiv 1/(e^x+1)$ and $x_i,x_j$ are the distances along
the two tubes.  This describes intertube tunnelling over a total
length $L$, but with the tunnelling rising smoothly over a lengthscale
$L_c$ from zero (far inside the ``leads'') to its maximal value.
Although this model for smooth contacts (\ref{eq:trirj}) does not
accurately mimic experimental situations, it provides a convenient way
in which to control the extent of scattering at the contacts. We shall
make use of this model (\ref{eq:trirj}) throughout the paper, choosing
sufficiently large $L_c$ to eliminate the effects of momentum
relaxation at the contacts and allow the bulk properties to be
studied.

The qualitative effects of $L$ and $L_c$ can be understood in simple
terms by considering the perturbative expression for conductance
(\ref{eqn:goldenrulecond1}) for a one-dimensional continuum
model with tunnelling amplitude $t(x_1,x_2) = t(x)\delta(x_1-x_2)$,
with
\begin{equation}
t(x)=t_0\left[\Theta
\left(x\right)-\Theta\left(x-L\right)\right]
\ast g_{L_c}\left(x\right)
\end{equation}
where the star denotes convolution, and
\begin{eqnarray}
g_{L_{c}}(x) &=&
\frac{1}{L_{c}}\frac{2\me^{-x/L_{c}}}{\left(1+\me^{-x/L_{c}}\right)^{3}}
\end{eqnarray}
For $L\gg L_c$ and $\delta \ll L_c$ this is an accurate approximation
to the functional form (\ref{eq:trirj}).  The matrix element of
the tunnel
coupling from states $k_{\rm A}$ to $k_{\rm B}$ is
\begin{eqnarray}
t_{k_{\rm A}k_{\rm B}}&\equiv & \frac{1}{L_N}\int\int t(x_{1},x_{2})\,\me^{-\mi k_{\rm B} x_{1}}\me^{\mi
  k_{\rm A} x_{2}}\,dx_{1}\,dx_{2}\\
& = &
\frac{t_{0}}{L_N}
\frac{2\sin{\left(\frac{\Delta k}{2}L\right)}}{\Delta k} \tilde{g}_{L_c}(\Delta k)
\label{eqn:sinc}
\end{eqnarray}
where $\Delta k \equiv k_{\rm B}-k_{\rm A}$ and
\begin{equation}
\tilde{g}_{L_c}(\Delta k) = \frac{\pi \Delta k
L_{c}(1+ i \Delta k L_{c})}{\sinh{\left(\pi\Delta k L_{c}\right)}}
\label{eqn:lcdependence}
\end{equation}
is the Fourier transform of $g_{L_c}(x)$, which was obtained by closing the
contour in the correct half-plane and summing the residues of an
infinite number of third order poles at $x=(2n+1) i \pi$.
This matrix element enters the perturbative expression for conductance
(\ref{eqn:goldenrulecond1}) to give
\begin{equation}
G= \frac{2e^2t_0^2}{\hbar^3 v_{\rm A} v_{\rm B}}
\frac{\sin{\left(\frac{\Delta k}{2} L\right)}^{2}}{\left(\Delta
k\right)^{2}} \left|\tilde{g}_{L_c}(\Delta k)\right|^2
\label{eq:osc}
\end{equation}

In the case of momentum conservation, $\Delta k \rightarrow 0$,
\begin{equation}
G = \frac{e^2 t_0^2}{2\hbar^3 v_{\rm A} v_{\rm B}} L^{2} \label{eqn:condlsq}
\end{equation}
That is, the conductance grows like
the {\it square} of the tunnelling length $L$, and is {\it independent}
of the contact length $L_c$: these results are consistent with the
conductance arising from the bulk of the system, unaffected by the contact regions.

For the situation of momentum non-conservation, $\Delta k \neq 0$ the
conductance (\ref{eq:osc}) is an oscillatory function of the total
length of the system, $L$.  This contribution arises from tunnelling
at the contacts, with the oscillatory dependence a consequence of
interference between the tunnelling at the contacts at two ends of the
sample.  Consistent with this explanation, the amplitude of this
oscillatory contribution to conductance depends sensitively on the
contact lengthscale $L_c$, with exponential suppression for large
$\Delta k L_c$.

\subsection{Rope Geometry}

\label{subsec:ropegeom}

We start by discussing the intertube conductance between the outer
shells of two neighbouring nanotubes that lie side by side in a
nanotube rope.  We choose an intertube separation of $b =
3.4\mbox{\AA}$ which is typical for a nanotube rope.

If the two tubes are identical, they are commensurate and momentum
conservation is satisfied at all energies. We therefore expect strong
coupling.  Indeed, we find in our numerical calculations (not shown)
that for two armchair tubes or two zigzag tubes the intertube
conductance is large, reaching of the order of $e^2/h$ even for very
short tunnelling regions between the tubes.

\label{subsec:incommens}

Typically, two tubes of different chirality have mismatched Fermi
momenta and are incommensurate. As discussed in
\S\ref{subsec:theoryincommens}, we expect tunnelling to be strongly
suppressed.
\begin{figure}
\centering
\includegraphics[scale=0.3]{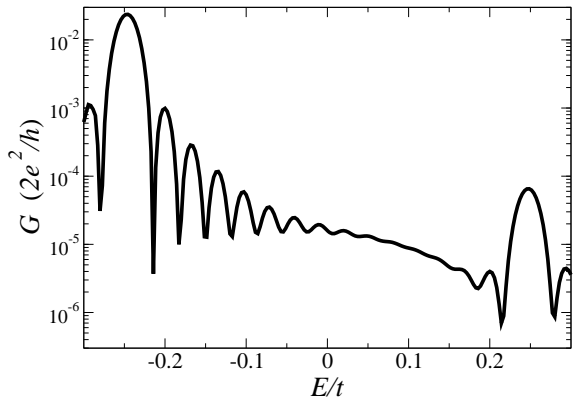}
\caption{Intertube conductance between $(10,10)$ and $(18,0)$ tubes,
  in the rope geometry with an intertube distance of
  $b=3.4\mbox{\AA}$. The total length of the \nt s is $L=87a$, and
  intertube tunnelling is introduced smoothly from zero over a length
  $L_{c}=2a$. The peaks at $E/t=\pm 0.25$ correspond to
  energy-momentum conserving processes described in the text.
}
\label{fig:clean}
\end{figure}
We investigate this effect for the case of $(10,10)$ armchair and
$(18,0)$ zigzag tubes. In Fig.\ref{fig:clean} we show the intertube
conductance as a function of the Fermi energy $E$. (We have suppressed
the effects of momentum relaxation at the contacts by choosing $L_c =
2a$.)  The conductance is very small for most energies.  However,
there are two peaks (with sidebands), at $E\approx \pm 0.25 t$.  At
these energies tunnelling can occur by the momentum-conserving
tunnelling process discussed at the end of
\S\ref{subsec:theoryincommens} in which a single reciprocal lattice
vector is added to the Fermi momenta of both armchair and zigzag
tubes. The presence and positions of these peaks are consistent with
the theory of Ref.~\onlinecite{maarouf} and formula
(\ref{eqn:maarouftkk}) which showed that the dominant tunnelling
processes involve the smallest number of \rlv s.  However, the heights
are not well reproduced by (\ref{eqn:maarouftkk}).  In particular, the
large asymmetry in the conductance at the peaks $E=\pm E_{0}$, by
several orders of magnitude, is not reproduced by
Eqn.~\ref{eqn:maarouftkk}.  These failings of
Eqn.~\ref{eqn:maarouftkk} show that one must go beyond this
approximate expression to achieve quantitatively accurate values for
the intertube conductance.

\begin{figure}
\centering
\includegraphics[scale=0.35]{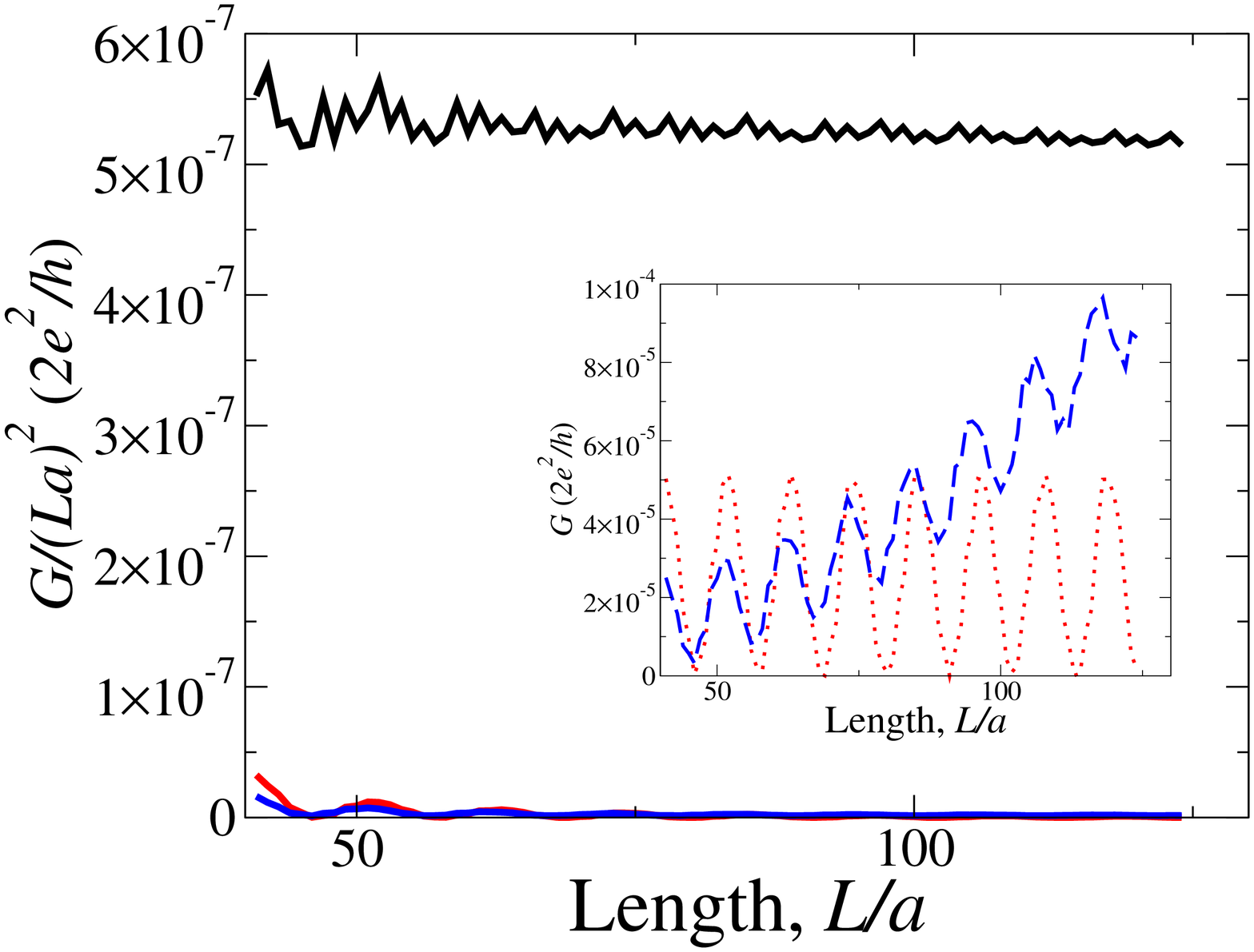}
\caption{(Color online) Length dependence of the intertube conductance for $(10,10)$ and
$(18,0)$ nanotube in a rope geometry a distance $b=3.4\mbox{\mbox{\AA}}$
apart. Results are shown as a function
of the total length, at $E=-E_0, 0, +E_0$ (solid, dotted, dashed,
respectively).  The main figure, which plots $G/(La)^2$, indicates clearly
that $G\sim L^2$ for $E=E_0$.  The inset, which plots $G$, indicates that
$G\sim L^2$ also for $E=-E_0$ but not for $E=0$.  These results are consistent
with a momentum conserving tunnelling processes at $E=\pm E_0$ but none at
$E=0$.  (A contact length of $L_{c}=2a$ has been chosen to minimise effects of
momentum relaxation at the contacts.)}
\label{fig:gvslen}
\end{figure}

\begin{figure}
\centering
\includegraphics[scale=0.33]{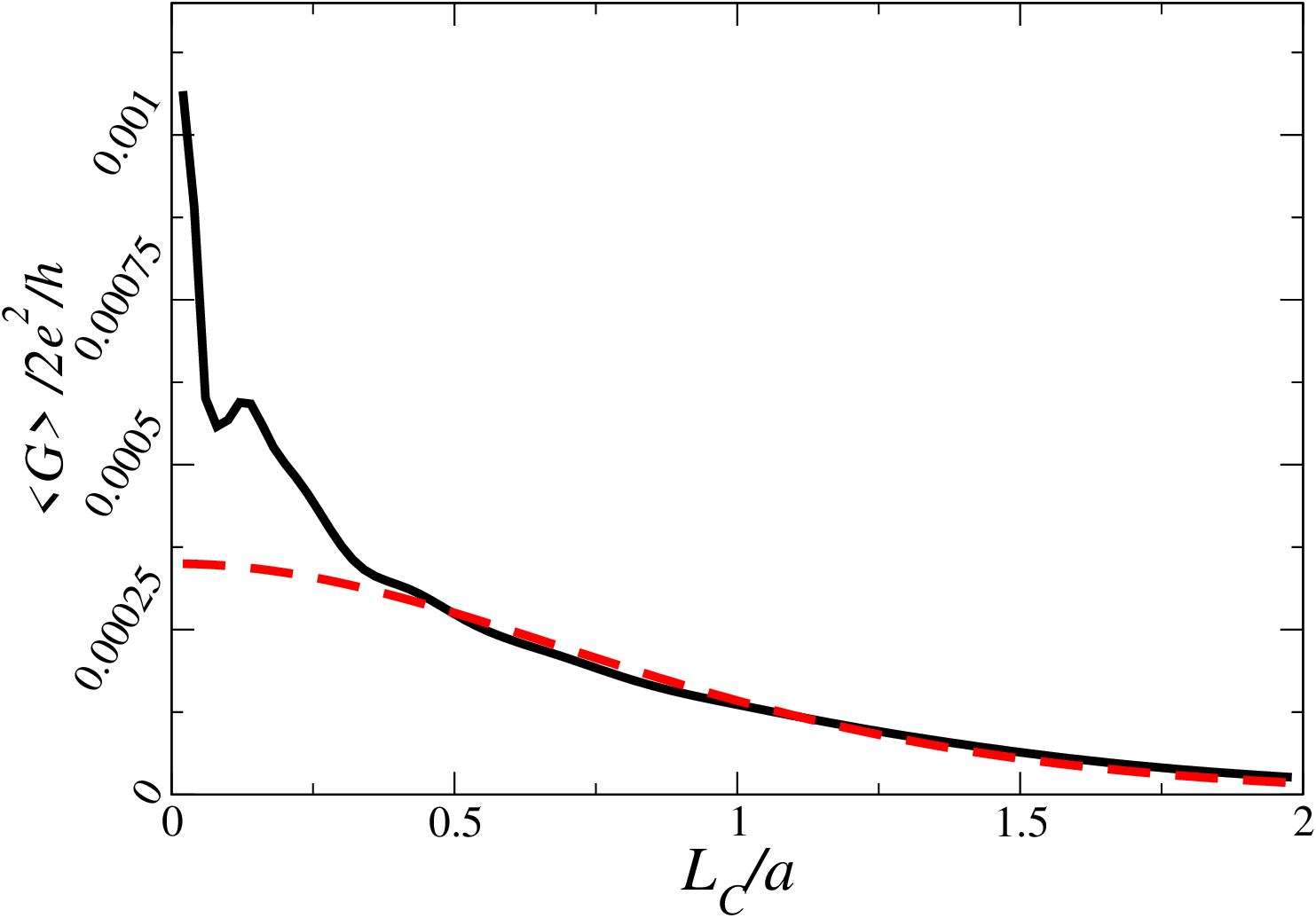}
\caption{(Color online) Mean intertube conductance at $E=0$ between $(10,10)$ and
  $(18,0)$ \nt s in the rope geometry a distance
  $b=3.4\mbox{\mbox{\AA}}$ apart, as a function of the contact length
  $L_{c}$.  The strong dependence on $L_c$ indicates that the
  conductance is dominated by momentum relaxation at the contacts.
  The fit (dashed line) is to the functional form of
  Eqn.~\ref{eq:osc}.
}
  \label{fig:avgsbs}
\end{figure}

The continuum model of \S\ref{subsec:contacts} accounts for the
qualitative features of the length-dependence of the conductance of
the $(10,10)$ to $(18,0)$ system. As illustrated in
Fig.~\ref{fig:gvslen}, at $E=\pm 0.25t$ we find that $G\propto L^2$,
characteristic of a bulk momentum-conserving tunnelling process [see
Eqn.~\ref{eqn:condlsq}].  At $E=0$ the conductance oscillates with
length, indicating that no momentum-conserving processes are
contributing.  The observed period of the oscillation is consistent
with the form $\sin^2 (\Delta kL/2)$, with the change in wavevector
$|\Delta k| = (2-\sqrt{3})2\pi/a$ expected from the difference between
the Fermi wavevectors of the armchair and zigzag tubes up to a shift
by \rlv s.\cite{footnote3}
Furthermore, the contact-induced contribution to the conductance is a
strong function of $L_c$.  In Fig.~\ref{fig:avgsbs} we show this
dependence, plotting the conductance at $E=0$ (averaged over two
periods of oscillation) as a function of contact length.
As shown in Fig.~\ref{fig:avgsbs} the rapid decay of conductance with
$L_c$ can be fit to the form given by Eqn.~\ref{eq:osc}. Deviations
from this form indicate the breakdown of the continuum approximation
for small $L_{c}\lesssim a$.

\subsection{Multiwall Geometry}

\label{subsec:mwsymtest}

For the multiwall geometry, in addition to momentum conservation along
the tube, the classification of the rotational symmetry,
\S\ref{sec:mwsym}, plays a key role.  In order to illustrate this we
have calculated the intertube conductance (\ref{eq:ginter}) for clean
DWNTs of the three different kinds: ``zero'' (3,3) in (15,0);
``intermediate'' (5,5) in (18,0); and ``strong'' (7,7) in
(21,0). These are chosen to have similar intershell spacing ($3.8
\mbox{\AA}, 3.7 \mbox{\AA}$ and $ 3.5 \mbox{\AA}$ respectively).  In
each case, since the tubes have mismatched Fermi momenta, the
conductance is small owing to momentum conservation along the tube.
However, the relative sizes of the conductances in these three cases
differ greatly: for the ``zero'' case, we find vanishing intertube
conductance to machine accuracy; the conductance of the ``strong''
coupling case is larger than that of the ``intermediate'' case by
about 8 orders of magnitude (results are not shown here). This very
significant difference in conductance indicates the importance of
distinction between ``strong'' and ``intermediate'' coupling cases.

\begin{figure}
\centering
\includegraphics[scale=0.33]{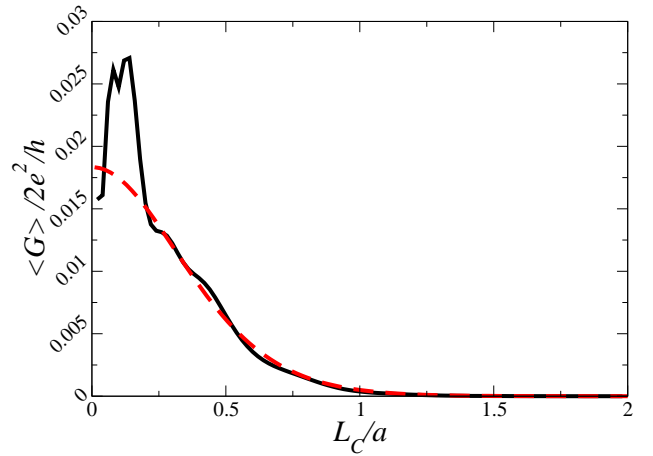} 
\caption{(Color online) Mean intertube conductance at $E=0$ between $(7,7)$ and
$(21,0)$ \nt s in a multiwall geometry as a function of the contact
length $L_c$.  The strong dependence on $L_c$ indicates that the
conductance is dominated by momentum relaxation at the contacts.
The fit (dashed line) is to the functional form
of Eqn.~\ref{eq:osc}.
}
\label{fig:avgmw}
\end{figure}

As in the rope geometry, the conductance is a strong function of the
contact length $L_c$. In Fig.\ref{fig:avgmw} this dependence is shown
for the average conductance at $E=0$ of the $(7,7)$ to $(21,0)$ DWNT
(which is the ``strong'' coupling case in terms of the rotational
symmetry).  Note that, for small values of the contact lengthscale, a
conductance of order $0.05 e^2/h$ can be induced. This is a sizable
conductance, given that it is due only to currents that flow in the
vicinity of the contacts. Our results emphasise that to understand the
intertube conductance between clean tubes, with mismatched Fermi
momenta and incommensurate periodicity, requires a very careful study
of the contacts: the conductance is dominated by the current flowing
in the vicinity of the contacts.

\subsection{Telescopic Junction}

\begin{figure}
\vspace{5pt}
 \centering
\includegraphics[scale=0.25]{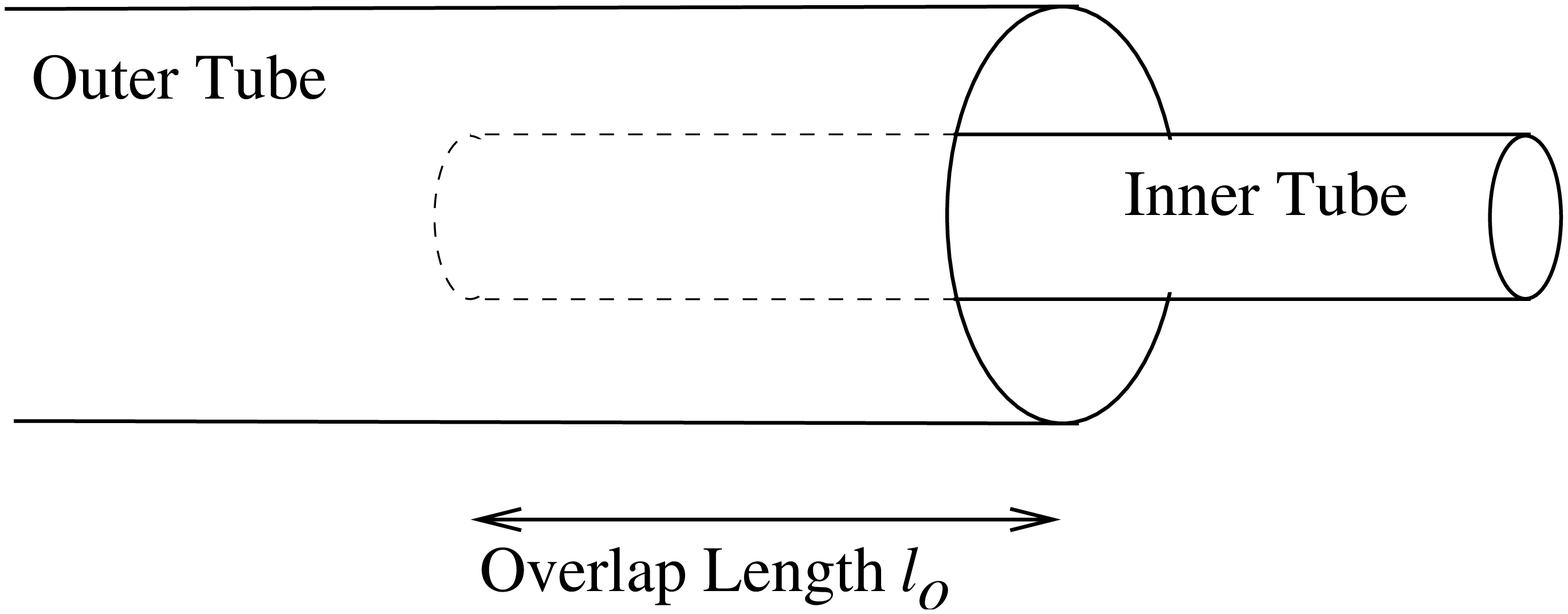}
\caption{Schematic picture of a telescopic junction.}
\label{fig:telesetup}
\end{figure}

\label{sec:telescopic} A particularly simple contact geometry
between two neighbouring shells of a MWNT or DWNT can be formed by
telescopic extension of the inner tubes\cite{cumings} and making
separate contacts to the inner and outer shells.\cite{cumings2}
Current that is passed between these two contacts is then forced to
flow between the two shells. (See Fig.\ref{fig:telesetup}.)  This
geometry involves a very abrupt change in the properties of the tubes
at the free ends in the inner and outer tubes.  We therefore
anticipate that the loss of momentum conservation at these edges will
play a significant role in the intertube conductance.

We have evaluated the conductance for telescopic junctions formed
from clean DWNTs.  To model this situation within our approach, the
leads $p=2,3$ in Fig.\ref{fig:sharpwires} are disconnected (the $\tau$
matrices discussed in \S\ref{sec:gf} are set to zero) and the
conductance between leads 1 and 4 is calculated.  Similar studies have
been performed previously for certain commensurate and incommensurate
tubes.\cite{kimchang,hansson,buia,tamura,yan} 

\begin{figure}
\centering
\includegraphics[scale=0.3]  {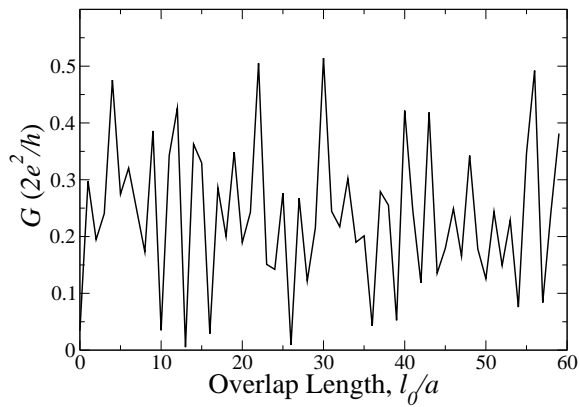}
\caption{Conductance of a telescopic junction between two
incommensurate $(6,6)$ to $(18,0)$ \nt s, as a function of the overlap
length, $\ell_o$. By our classification of the rotational symmetry the
coupling is ``strong''.  The two tubes are strongly coupled even for
small overlap lengths.}
\label{fig:teleincom6to18}
\end{figure}

\begin{figure}
\centering
\includegraphics[scale=0.3]{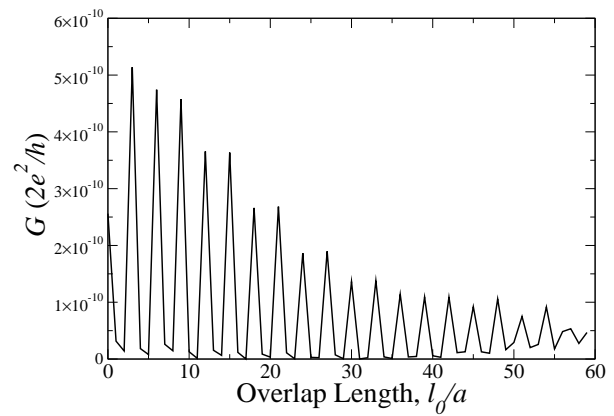}
\label{fig:teleincom6}
\caption{Conductance of incommensurate $(10,10)$ to $(9,0)$ \nt s for
a telescopic junction, as a function of the overlap length $\ell_o$.
By our classification of the rotational symmetry the coupling is
``intermediate''.  The two tubes remain weakly coupled even for large
overlap lengths.}
\label{fig:teleincom10to9}
\end{figure}

Here we focus on the case of incommensurate tubes with mismatched
Fermi momenta, and discuss the influence of the rotational symmetry
classification of \S\ref{sec:mwsym}.
A telescopic junction does not break the rotational symmetry, so this
classification remains relevant. We present results for both
``intermediate'' [$(10,10)$ to $(9,0)$] and ``strong'' coupling
[$(6,6)$ to $(18,0)$] in Figs.~\ref{fig:teleincom6to18} and
\ref{fig:teleincom10to9}. (For the ``zero'' case, we find that the
conductance vanishes, as expected by the symmetry analysis.)
For the ``strong'' coupling, Fig.~\ref{fig:teleincom6to18}, we indeed
find very efficient coupling.
For even small overlap lengths of a few lattice constants, the
conductance is of order $e^2/h$. Attempting to describe the system as
two standing waves in different tubes is no longer valid in this
regime: since the intershell conductance is of order $e^2/h$ the tubes
are strongly coupled and perturbation theory is inaccurate. For
intermediate coupling, Fig.\ref{fig:teleincom10to9}, we find an
oscillatory intershell conductance but of a very much smaller
magnitude. Increasing the overlap does not increase the conductance,
consistent with the fact that this is a system with mismatched Fermi
momenta, for which there are no bulk momentum-conserving tunnelling
processes.
These results again highlight the striking difference between
rotational symmetry classification as ``strong'' and
``intermediate''. In the former case, the relaxation of momentum at
the junction between SWNT and DWNT region is enough to cause a
conductance of order $e^2/h$; in the latter, the intershell
conductance is many orders of magnitude smaller.

\section{Effects of Defects on Intertube Conductance}

\label{sec:defects}

In \S\ref{sec:clean} the intertube conductance was studied for clean
(defect-free) tubes.  Momentum conservation was shown to play a very
important role in the intertube conductance between tubes with
mismatched Fermi momenta.
In real samples, electrical conductivity typically depends vitally on
the number and type of defects present.  One can expect defects to act
as scattering centres and to relax momentum conservation. This general
point was noted in Ref.~\onlinecite{maarouf}. However, we are not
aware of any studies taking into account the effects of defects on
intertube or intershell transport. In this section we provide
calculations of the effects of defects of different types on the
conductance of incommensurate nanotube ropes and MWNTs. In particular,
we discuss the effects of individual scatterers (vacancies and other
localised defects) on intertube transport in cases where the clean
tubes would have small intertube conductance.  We shall then discuss
how these results can be extended to a finite density of scatterers.

In all the studies we report below, we use a contact length of
$L_{c}=2a$ to suppress the effect of momentum relaxation at the
contacts and the total length of the central region is made
sufficiently long (of order $L=100a$) that we are able to extract the
influence of the impurity in the bulk system without finite size
effects.

\subsection{Vacancies}

\label{sec:singlevac}

While rare in natively grown nanotubes, vacancies can be introduced by
sputtering.\cite{stahl} Here we determine the effects of vacancies on
intertube conductance in ropes and MWNTs.

\subsubsection{Rope Geometry}

\begin{figure}
\centering
\includegraphics[scale=0.35]{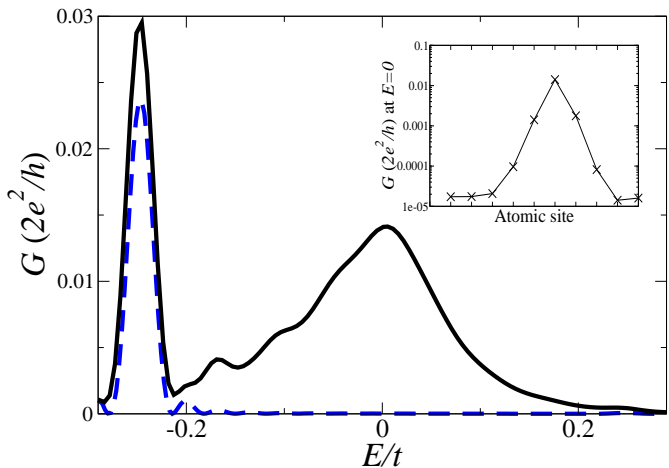}
\caption{(Color online) Intertube conductance between a $(10,10)$ armchair and
  $(18,0)$ zigzag in the rope geometry a distance $b=3.4\mbox{\AA}$
  apart. Results are shown both for clean \nt s (dashed line) and with
  a single vacancy placed on a site halfway along the armchair tube
  (solid line).  The inset shows the dependence of the conductance at
  $E=0$ on the location of the vacancy at the atomic sites around the
  circumference of the armchair tube.}
\label{fig:1010to180gvsE}
\end{figure}

In Fig.\ref{fig:1010to180gvsE} we present results for the effects of a
vacancy on the conductance between $(10,10)$ and $(18,0)$ \nt s in the
rope geometry.  The nanotubes both have diameters of about $1.4$nm and
are placed $3.4\mbox{\mbox{\AA}}$ apart. The vacancy, which we
introduce in the same way as in \S\ref{sec:singletube}, is
placed on the armchair tube at the point of closest approach.

In Fig.~\ref{fig:1010to180gvsE} the dashed line shows the intertube
conductance for clean tubes, which was discussed in
\S\ref{subsec:incommens} and shown in Fig.\ref{fig:clean} (on a
logarithmic scale).  With the introduction of a vacancy the
conductance around $E=0$ is dramatically increased to a value of order
$G\sim 0.01 (2e^2/h)$.  Note that this is the conductance induced by a
{\it single} vacancy.  In a large system containing many such defects
one can expect even larger enhancements of the intertube
conductance. If the defects contribute incoherently (see
Sec.\ref{sec:incomain}), the intertube conductance will increase
proportionately to the number of defects.

We have investigated the dependence on intertube conductance of the
position of the vacancy. (The effect of varying the strength of the
potential will be investigated in the next section.)  The inset to
Fig.~\ref{fig:1010to180gvsE} plots the results obtained for intertube
conductance at $E=0$ by moving the vacancy around the circumference of
the tube on the $10$ sites of the $(10,10)$ armchair tube that are
crystallographically equivalent.  The position of the vacancy has
significant impact on the intertube conductance, with the largest
enhancement of intertube conductance when the vacancy site is closest
to the other nanotube.  This indicates that the wave that is
backscattered by the vacancy in the defective tube has a large
amplitude localised in the vicinity of the vacancy.

An experimental investigation of the effects of vacancies (introduced
intentionally by sputtering) on the conductance of nanotube ropes was
carried out by Stahl \emph{et al.}.\cite{stahl} The results were
interpreted in terms of the intertube tunnelling between metallic tubes
of {\it matching} chirality in a rope consisting of a random
distribution of tubes of random chiralities.  Tunnelling between \nt s
of different chiralities was ignored in the analysis, based on the
existing theory\cite{maarouf} for clean tubes which showed that this
tunnelling is heavily suppressed as a result of momentum conservation.
Our results show that, owing to the loss of momentum conservation in
the vicinity of the vacancy, the intertube transport between nearest
neighbour tubes of {\it different} chirality can occur
efficiently. Even for a single vacancy, the intertube conductance
between two neighbouring tubes can be as large as $0.01 (2e^2/h)$,
corresponding to a resistance of around $1 \mbox{M}\Omega $. This is
well within the range of intertube resistances that was measured in
Ref.~\onlinecite{stahl}.  Our results therefore indicate that owing to
the loss of momentum conservation at the vacancies, the tunnelling
between incommensurate certainly \nt s cannot be neglected in these
experiments, and is likely to play a significant role.

\subsubsection{Multiwall Geometry}

\begin{figure}
\includegraphics[scale=0.35]{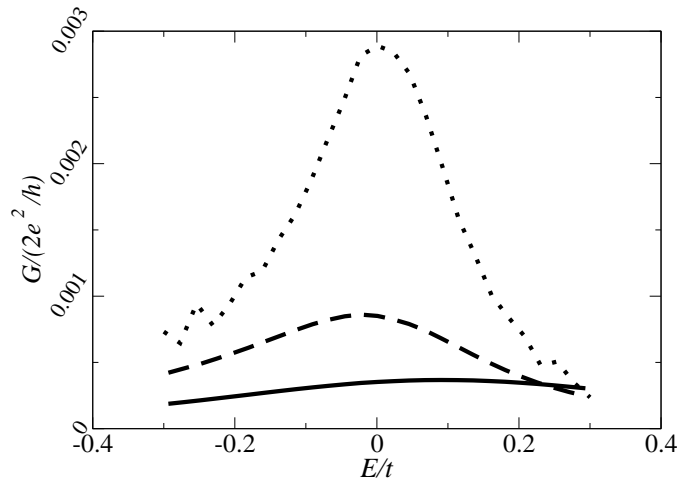}
\caption{Intershell conductances for DWNTs in the presence of a single
vacancy (placed on the inner tube), for the three types of coupling
determined by the angular symmetry of \S\ref{sec:mwsym}. The solid
line is  a $(3,3)$ to $(15,0)$ system (``zero'' coupling);
the dashed line is a $(5,5)$ to $(18,0)$ system (``intermediate'' coupling);
and the dotted line is a $(7,7)$ to $(21,0)$ (``strong'' coupling).}
\label{fig:mwvac}
\end{figure}

We now consider the effects of a vacancy in a DWNT geometry.  The
vacancy breaks the rotational symmetry described in
\S\ref{sec:mwsym} thereby removing the distinction between the
different strengths of coupling of the metallic subbands.  Therefore,
if all other parameters are equal, one would expect the differences
between ``Zero'', ``Intermediate'' and ``Strong'' cases to disappear.
Fig.~\ref{fig:mwvac} shows the conductance for the same three DWNTs
considered in \S\ref{subsec:mwsymtest}, now with a single vacancy in the
armchair tube.  The magnitude of the intertube conductance for all
three types is now $G\approx 10^{-3}\left(2e^{2}/{h}\right)$ around
$E=0$. Recalling that the intertube conductances for clean tubes,
discussed in \S\ref{subsec:mwsymtest}, differ by many orders of
magnitude in these three cases, one indeed see that the vacancy does
break the distinction between these cases giving conductances of
similar magnitudes. Furthermore, the overall size of the conductance
is very much increased.  Again, the presence of even a single vacancy
can very significantly affect the conductance between the shells of a
MWNT.

\subsection{Other Types of Defect}

\label{sec:mwdefects}

The nature of the dominant defect scattering in high quality nanotubes
is not well characterised. Indeed, it is likely to vary between
systems involving different preparation methods and different
geometries.  The disparity between the intershell MWNT conductance
experiments of vertically suspended MWNTs\cite{frank} and those where
the MWNTs are placed on a substrate\cite{bourlon} suggests that
defects may arise due to substrate effects.  We have studied the
effects of various types of localised defects, including 
potential defects and deformations of the outer tube arising from
substrate roughness.

We focus on the case of MWNTs, in order to make connection with the
experimental measurements of Bourlon {\it et al.}\cite{bourlon} We
have calculated the effects on the conductance at $E=0$ of a single
defect in the outer shell of a DWNT, consisting of a $(13,13)$
armchair (outer shell) and a $(12,0)$ zigzag (inner shell) \nt
. According to our classification of rotational symmetry, this
combination is of ``intermediate'' coupling, which is the typical
case.
The effect of a single defect is determined by choosing a sufficiently long section of tube
$L$ and sufficiently smooth contacts $L_c$ that finite size effects
are negligible.  In order to quantify the effect of the scattering, we
report dimensionless measures of intratube resistance, and intertube
conductance, defining the quantities
\begin{eqnarray}
R_s\equiv 1-\frac{T_{s}}{2}&=&1-\frac{1}{2}\sum_{\mu,\mu'}T_{\mu,1;\mu',2}
\label{eqn:singscattrans1} \\
I_s&\equiv &\sum_{\mu,\mu'; p=1,2; p'=3,4}T_{\mu,p;\mu',p'}
\label{eqn:singscattrans2}
\end{eqnarray}
$R_s$ is a measure of the intratube backscattering,\cite{footnote4}
and $I_s$ is a measure of the intertube conductance.  In
\S\ref{subsec:expt} we shall discuss how these quantities can be
used to compare with experiment.\cite{bourlon}

\subsubsection{Potential Scatterer}
\label{sec:singdefpot}

We consider a potential centred on a site $\bm{r}_0$ of the outer
\nt~with strength $V_0$ and correlation length, $\xi$, such that there
is an additional site potential on this tube of
\begin{equation}
V_i =V_0 \,\me^{-{\modulus{\bm{r}_{i}-\bm{r}_{0}}^2}/{\xi^2
}}
\label{eqn:corrpot}
\end{equation}
This is a model for several types of defect: molecules absorbed onto
the surface of the MWNT; a residual potential due to the close
proximity to the substrate; nitrogen substitution, which can be
modelled by choosing $V_{0}\approx-2.5\mbox{eV}$.\cite{kosty}

\begin{figure}
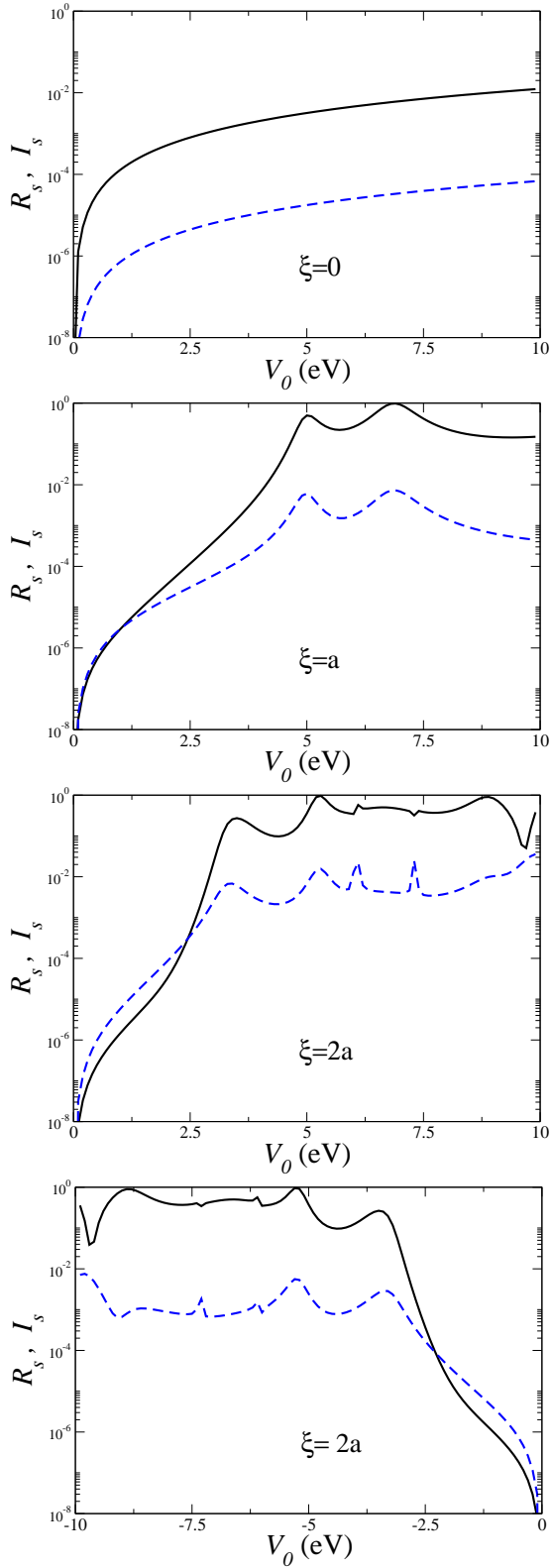

\centering
{\includegraphics[scale=0.29]{tc_fig13a.eps}}
{\includegraphics[scale=0.29]{tc_fig13b.eps}}
{\includegraphics[scale=0.29]{tc_fig13c.eps}}
{\includegraphics[scale=0.29]{tc_fig13d.eps}}
\caption{(Color online) The dimensionless measures of backscattering, $R_s$, (solid
line) and intertube scattering, $I_s$, (dashed line) for a (13,13) to
(12,0) DWNT with a single potential defect, of varying strengths
$V_{0}$ and correlation length $\xi$ [see Eqn.~\ref{eqn:corrpot}]. The
total length of the system is $L=87a$ with tunnelling introduced
smoothly via a contact length of $L_{c}=2a$, so boundary effects are
minimal and the results are characteristic of a single defect in the
bulk.}
\label{fig:mwpotentialsing}
\end{figure}
In Fig.~\ref{fig:mwpotentialsing} the dimensionless parameters of
Eqns.~\ref{eqn:singscattrans1} and~\ref{eqn:singscattrans2} are
plotted against potential strength $V_{0}$ for different values of the
correlation length, $\xi$. In each case, the backscattering $R_s$ and
intertube coupling $I_s$ increase at small $V_0$ as $V_0^2$ consistent
with expectations from perturbation theory.  In this weak scattering
regime, the relative sizes of the backscattering and intertube
scattering depend strongly on the correlation length $\xi$.  One might
expect that increasing the range of the potential $\xi$ at fixed
amplitude $V_0$ would cause a larger effect of the defect as the total
integrated weight of the potential increases.  This is true for $I_s$
and for $R_s$ at large $V_0$. However, for small $V_0$, the intratube
backscattering $R_s$ actually falls with increasing $\xi$, even to
below $I_{s}$ for $\xi=2.0a$: in this case, the defect more
efficiently causes intertube scattering than it does intratube
backscattering.  This strong dependence on the correlation length can
be understood as a consequence of the suppression of intratube backscattering
due to the conservation of ``pseudo-spin'' by the
(perturbative) scattering from a potential with range larger than a
lattice constant.\cite{andohelicity1,andohelicity2,mceuen} For large
$\xi$ the backscattering parameter $R_s$ becomes small; there is no
such suppression of intertube scattering, so $I_s$ remains large.

The effects of the potential scatterer are very similar for both
positive and negative $V_0$.  For $\xi=0$, both $R_{s}$ and $I_{s}$
are symmetric in $V_{0}$. For $\xi\neq 0$, an asymmetry in $V_0$
arises, but, as shown in Fig.~\ref{fig:mwpotentialsing}, even for
$\xi=2a$ the asymmetry remains very small.

\subsubsection{``Dent'' Scatterer}
\label{sec:singdisdef}
One possible source of scattering in the measurements of Bourlon
\emph{et al.}~\cite{bourlon} is the roughness of the substrate, which
could lead to deformation of the MWNT.\cite{bourlon} We investigate
this by considering a ``dent'' in the outer shell, which when placed
at a position $\bm{r}_0$ gives rise to a correlated displacement of
the atoms. We displace the atom $i$ (with undeformed position
$\bm{r}_i$) in the radial direction towards the tube axis by a
distance
\begin{equation}
\delta r_{i} = d_0\; \me^{-{\modulus{\bm{r}_{i}-\bm{r}_{0}}^2}/{\xi^2 }}
\, .
\label{eq:dent}
\end{equation}
These displacements will change the inter- and intra-tube hopping
parameters and could also lead to additional onsite
potentials. Provided the curvature of the tube is negligible, the
correction to the intratube hopping and onsite potential is of order
$\left({d_0}/{a}\right)^2$ (by symmetry there can be no term linear in
$d_0$). However, the correction to the intertube hopping is of order
${d_0}/{a}$. We therefore consider only the changes to the intertube
hopping, which is accurate for small deformations $d_0 \ll a$.

\begin{figure}
\centering
\vspace{17pt}
\includegraphics[scale=0.35]{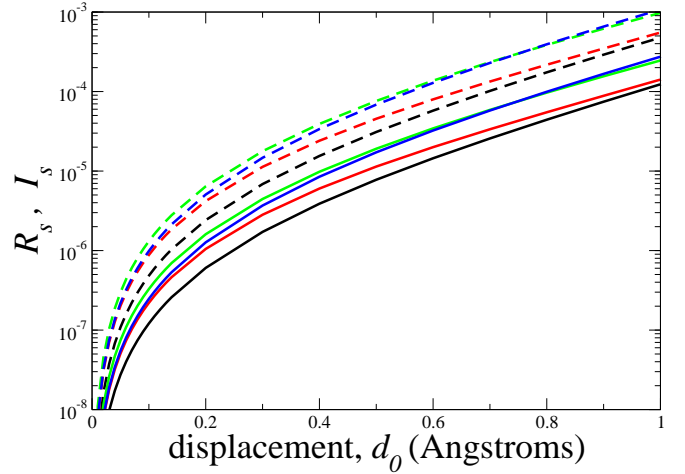}
\caption{(Color online) The dimensionless measures of backscattering, $R_s$, (solid
lines) and intertube scattering, $I_s$, (dashed lines) for a (13,13)
to (12,0) DWNT with a single ``dent'' in the outer shell as a function
of the displacement $d_0$ and correlation length $\xi$ [see
Eqn.~\ref{eq:dent}]. Results are shown for the set of correlation
lengths $\xi = 0.00, 1.76, 3.53, 5.29\mbox{\AA}$, which may be
assigned to the graphs by noting that both $R_s$ and $I_s$ increase
with increasing $\xi$ for $d_0 \sim 1\mbox{\AA}$.}
\label{fig:singmwsdentsmall}
\end{figure}

Fig.~\ref{fig:singmwsdentsmall} shows the results of calculations of
$R_s$ and $I_s$, Eqns.~\ref{eqn:singscattrans1}
and~\ref{eqn:singscattrans2}, for a single dent scatterer of various correlation lengths $\xi$ as a function of the strength $V_0$.
Compared to the potential scatterers, the ``dent'' scatterers give
qualitatively different results.  Now, the intertube conductivity is
always higher than the intratube resistivity.  This is a consequence
of the displacement having no effect on the intratube tight-binding
parameters (valid for $d_0 \ll a$): intratube backscattering involves
a second order process in which the electron tunnels to and from the
other tube.  Thus our results indicate that the dent scatterer
especially enhances the intertube conductance as compared to the
intratube resistance.

\subsubsection{Experimental Comparison}

\label{subsec:expt}

In the experimental studies of Bourlon \emph{et al.}\cite{bourlon},
non-local voltages induced between contacts to the outer shell of
MWNTs were used to extract the intrashell resistivity,
$\rho^{\text{exp}}_{\rm intra} =\left(6-25\right) \mbox{k}\Omega \mu
\mbox{m}^{-1}$, for each of the outer two shells and an intershell
conductivity, $g^{\text{exp}}_{\rm inter} =\left[\left(3.7-20\right)
\mbox{k}\Omega\right]^{-1}\mu \mbox{m}^{-1}$, between these two outer
shells. (Contributions from other shells were shown to be small.)
As emphasised in Ref.~\onlinecite{bourlon}, the intertube conductivity
cannot be understood within existing theoretical descriptions of
intershell coupling in incommensurate nanotubes. We believe, as
suggested in Ref.~\onlinecite{bourlon}, that the discrepancy is due to
the effects of disorder which was not considered in previous
theoretical works. We can use our results on impurity scattering to
estimate the intrashell resistivity and intershell conductivity in the
experiments.  Since the nature and density of defects are very
uncertain in these experimental systems, our approach is to search for
parameter regimes that are consistent with the experimental
results. This provides constraints on the nature of the defects in
these systems.

\label{sec:incomain}

Comparing the results of our calculations with these experimental
results\cite{bourlon} requires a number of simplifying assumptions.

While the MWNTs used in the experiments contained around $20$ shells,
since transport was shown to be largely confined to the outer two
shells we consider just two coaxial shells.  We use the results shown
above on the $(13,13)$ to $(10,0)$ system, which is a large radius
system showing the effects of incommensurability, and with typical
``intermediate'' coupling.  The overall diameter of this system, of
about $d=16\mbox{\AA}$, is much smaller than the diameter
$D=17\mbox{nm}$ of the outer shell of the MWNTs in the experiment. We
take account of this difference by scaling the scattering rates we
obtain in the numerics on a system with diameter $d$ by a factor of
$(d/D)^2 \sim 1/100$.\cite{white} [This reduction reflects the
reduction in the matrix element of a localised potential as $1/D$, and
thus a reduction in scattering rate (using Fermi's golden rule) as
$1/D^2$.]

In the experimental system the tubes are doped such that several modes
are occupied, with the estimated number of modes $N_{\rm
m}=10-20$.\cite{bourlon} There are then $2N_{\rm m}$ Fermi points in
each shell. Generically, the Fermi momenta of adjacent shells will not
coincide and the shells will be incommensurate.
We take the results of the above calculation of the
effects of defects on the transport between the metallic bands of
armchair and zigzag tubes as a measure of the scattering probabilities
between states at the individual Fermi points of the two outer shells
of the MWNTs in Ref.~\onlinecite{bourlon}.
We consider the limit of weak backscattering and weak intershell
scattering from a single defect.
In our numerical results, each shell has $4$ Fermi points, and we use
the total intertube scattering probability $I_s$ to obtain the
intertube scattering probability between individual Fermi points as
$I_s/4^2$ (assuming that the 4 Fermi points in one tube scatter
equally to the 4 Fermi points in the other).  Then, for $N_{\rm m}$
modes, we extend to $2N_{\rm m}$ Fermi points in total to obtain that
the scattering probability for a single incoming mode into any of the
modes on the second tube is
\begin{equation}
p_I = (2N_{\rm m}) \frac{I_s}{4^2} \left(\frac{d}{D}\right)^2
\end{equation}
which incorporates the reduction of $({d}/{D})^2$
discussed above.
Similarly for the reflection probability, extending from the case (in
the numerics) of two backscattering modes to $N_{\rm m}$ backscattering
modes,  the mean backscattering probability for a single
incoming mode into any of the modes on the second tube is
\begin{equation}
p_R = (N_{\rm m}) \frac{R_s}{2} \left(\frac{d}{D}\right)^2 \, .
\end{equation}
We have ignored current conservation in these estimates, which would
involve the introduction of compensation between intertube and
backscattering probabilities; this assumption is valid provided
scattering is weak, and both $p_I$ and $p_R$ are small.

The experiments involve the scattering by many defects, while the
above numerical results are for a single scatterer.  We extend to
larger numbers of scatterers by assuming that the scatterers combine
{\it incoherently}, with $n_{\rm s}$ defects per unit length.  This
will give a good approximation to the transport properties provided
the system is not in a strongly localised regime. The experimental
samples are far from the strongly localised regime even down to low
temperatures: from the experimental measurements the conductance on
the lengthscale of the contact separation ($\lesssim 1\mu\mbox{m}$)
remains large compared to $e^2/h$.

Within these assumptions, and using the analysis presented in
Appendix~\ref{app:inco}, we find that, for $n_{\rm s}$ scatterers per
unit length, the intertube conductivity $g_{\rm inter}$ and intratube
resistivity $\rho_{\rm intra}$ (both per unit length) are
\begin{eqnarray}
g_{\rm inter} \equiv \frac{G_{\rm inter}}{L} & = &
\left(\frac{2e^2}{h}\right) \frac{I_s   N_{\rm m}^2}{4} n_{\rm s} \left(\frac{d}{D}\right)^2 \\
\rho_{\rm intra} \equiv \frac{R_{\rm intra}}{L} & = &
\left(\frac{h}{2e^2}\right)  \frac{R_s}{2} n_{\rm s} \left(\frac{d}{D}\right)^2
\label{eq:rho}
\end{eqnarray}
Note that, within this weak scattering approximation, the resistivity is
independent of the number of modes: as $N_{\rm m}$ increases, the increased
number of Fermi points into which backscattering can occur (increasing the
resistivity) is compensated by the increase in the number of modes available
for conduction (increasing the conductance).  On the other hand the intertube
conductance increases as the \emph{square} of the number of modes, $N_{\rm
m}$: one factor coming from the number of incoming modes, and one factor from
the increased number of Fermi points into which any one mode can be scattered
in the other shell.

We are now in a position to compare our results with the experimental
results.\cite{bourlon} It is helpful to note that within our theory
the ratio of $g_{\rm inter}$ to $\rho_{\rm intra}$ is independent of
the defect density
\begin{equation}
\frac{g_{\rm inter}}{\rho_{\rm intra}} = \left(\frac{2e^2}{h}\right)^2 \frac{N_{\rm m}^2}{2} \frac{I_s}{R_s}
\label{eq:ratio}
\end{equation}
In the experiments,\cite{bourlon} this ratio is of order $g^{\rm
exp}_{\rm inter}/\rho^{\rm exp}_{\rm intra} \sim
\left({2e^2}/{h}\right)^2$. Agreement of the theory (\ref{eq:ratio})
with this experimental value requires $R_s/I_s \sim 0.5 N_{\rm m}^2
\sim (50-200)$, using the estimate\cite{bourlon} $N_{\rm m}\sim
10-20$.  Within the wide range of uncertainty in the parameters, the
experimental results are consistent with our results for potential
defects with vanishing correlation length $\xi = 0$, for which we find
$R_s/I_s \simeq 180$ over the range of $V_0$ shown in
Fig.\ref{fig:mwpotentialsing}(a), or for potentials with a small
correlation length.  In our calculations, the ratio $R_s/I_s$ becomes
small for potential scatterers with large correlation lengths; this
reduction is likely to be overestimated in a calculation based on
metallic bands alone.  Note that in the case of ``dent'' defects we
find that $R_s/I_s \ll 1$ in all cases
(Fig.\ref{fig:singmwsdentsmall}).  Thus, our analysis shows that the
experimental results are consistent with the dominant disorder being
potential scatterers with a small correlation length, but are {\it
not} consistent with the dominant disorder being ``dents'' in the
outer tube caused by substrate roughness.\cite{dents}

To estimate the strength of the disorder, we focus on the case of
potential defects with vanishing correlation length, $\xi=0$.  From
the results in Fig.\ref{fig:mwpotentialsing}(a), we find that $R_s
\sim 1.3 \times 10^{-4} \left(V_0/\mbox{eV}\right)^2$ over the range
of $V_0$ shown.  Matching (\ref{eq:rho}) to $\rho^{\rm exp}_{\rm
intra}\sim 10 \mbox{k}\Omega \mu \mbox{m}^{-1}$ sets an estimate for
the required number of potential scatterers $n_{\rm s}$ per unit
length as a function of their strength $V_0$. Converting to the number
of scatterers per atom, $\sqrt{3} a^2 n_{\rm s}/(4\pi D)$, we find
agreement for a root mean square potential energy fluctuation on each
site of about $0.8 \mbox{eV}$.  This is comparable to the disorder
strengths required by other theoretical studies to account for a
resistivity of this size.  [Indeed, Eqn.~\ref{eq:rho} closely matches
the results of Ref.~\onlinecite{white}, for $N_m=2$, relating
mean-free-path to resistivity by the Drude formula.]

In summary, our theory shows that the dramatic increase of intershell
coupling arising from scattering from disorder is sufficient to
account for the experimental observations. We find quantitative
agreement (within the uncertainties of the experimental parameters)
for a model of disorder consisting of short-range potential
scatterers.  We note that our theory shows that the relative sizes of
the intershell scattering and intrashell backscattering depend
strongly on the number of modes $N_{\rm m}$, with $g_{\rm
inter}/\rho_{\rm intra}\propto N_{\rm m}^2$.  It would be interesting
to test this prediction in experiment.

\section{Conclusions}

\label{sec:conclude}

We have studied the transport between aligned carbon nanotubes (or
nanotube shells) in both rope and multiwall geometries. We focused on
cases of two metallic nanotubes of different chiralities, with
mismatched Fermi momenta and incommensurate periodicities.  Our
numerical calculations show clear evidence of the importance of
momentum conservation, and of a classification we introduced for the
rotational symmetry of a pair of shells in a MWNT. For clean tubes, we
find that the loss of momentum conservation at the contacts dominates
the intertube transport.

We studied in detail the effects of localised defects on the intertube
transport in incommensurate nanotube ropes and MWNTs.  Defects break
momentum conservation, and the rotational symmetry in multiwall
geometries. We find that the intertube conductance of incommensurate
nanotubes is very dramatically increased by even a single vacancy. Our
results show that the introduction of vacancies to nanotube ropes
enhances the tunnelling between incommensurate tubes sufficiently to
make this defect-mediated tunnelling an important transport
mechanism.\cite{stahl} We calculated the effects of localised defects
of various types on the intra- and inter-shell transport of MWNTs. For
a potential scatterer, we find a strong dependence on the range,
$\xi$, of the potential: for large $\xi$, intratube backscattering is
suppressed\cite{andohelicity1,andohelicity2,mceuen} but intertube
scattering remains large. For a ``dent'' scatterer, involving a
localised depression of the outer shell, intertube conductance is very
much enhanced with intratube backscattering remaining small.

Our results establish the extreme sensitivity to disorder of the
intertube transport of incommensurate nanotubes with mismatched Fermi
momenta. They provide the first quantitative description of the 
effects of this scattering on the transport between nanotube shells in
rope and multiwall geometries. A comparison of the results of our
calculations with experimental measurements\cite{bourlon} shows
agreement with our theory for a model of disorder consisting of
short-range potential scatterers. Potential scatterers with large
correlation lengths and deformations of the outer shell are not
consistent with the experimental results.

Finally, we would like to comment briefly on the effects of
electron-electron interactions.  In a one-dimensional metal,
interactions lead to a smearing of the Fermi surface, so that there is
no longer a discontinuity in the occupation numbers at the Fermi
surface.\cite{schonreview} Owing to the central role played by
momentum conservation in our considerations, it is important to ask
whether this smearing of the Fermi surface could affect our
results. We find that including interactions within a Luttinger liquid
analysis does not change the essential behaviour described above for
non-interacting electrons:\cite{thesis} the intertube conductance is
still strongly suppressed for incommensurate nanotubes with mismatched
Fermi wavevectors, and is strongly enhanced by impurity scattering.
The interactions will, in addition, cause the overall intertube
conductance to be suppressed as a power law in temperature (or bias),
as is familiar from tunnelling into a Luttinger liquid. These
interaction effects will be weaker at high temperatures and for MWNTs
with many occupied subbands.\cite{egger}

\acknowledgments We are grateful to Dr R. Roemer and Prof. M. Payne
for helpful comments.  This work was supported by the EPSRC Grant
GR/R99027/01.

\section{Green's functions for semi-infinite nanotubes}
\label{app:ntgreenfns}

To determine the Green's functions  we make use of the form
\begin{equation}
\left<i,\alpha \right| G \left| j,\beta
\right>=\sum_{\mu ,k}\frac{\left<i,\alpha
|\Psi_{\mu ,k}\right>\left<\Psi_{\mu ,k}|j,\beta
\right>}{E-\epsilon_{\mu ,k}+i\delta} \label{eqn:gfsumdefapp},
\end{equation}
where $|\Psi_{\mu ,k}\rangle$ are energy eigenstates, $\mu$ is the mode
index and $k$ is the wavevector along the axis of the \nt.  Positions
are labelled by $i,\alpha$, with $i$ labelling the 1D unit cell (along
the tube axis), and $\alpha$ the atoms within this 1D cell.

\subsection{Armchair Tube}

The metallic bands in an armchair \nt~have energy dispersion
relations
\begin{equation}
\epsilon_{\mu=\pm,k}=\mp t\left[1-2\cos{\left(\frac{ka}{2}\right)}\right]
\label{eq:ac}
\end{equation}
where $\mu= \pm$ labels the two metallic bands, $k$ is wavevector along the
axis of the nanotube.  For a tube with periodic boundary conditions,
we write the wavefunction $\ket{\Psi_{\mu ,k}}$ in
the form
\begin{equation}
\langle i,\alpha| \Psi_{\mu ,k}\rangle =
\frac{\me^{\mi kx_{i}}}{\sqrt{\frac{N_{\rm 1D}}{2}}}\chi_{\mu ,k}(\alpha)
\label{eqn:ntwf}
\end{equation}
where $x_{i}$ is the distance of the 1D unit cell along the tube axis,
$N_{\rm 1D}$ is the number of 1D unit cells.  The functions
$\chi_{\mu,k}$ are easily obtained from the free-particle eigenstates;
explicitly for an $(n,n)$ tube they are
\begin{equation}
\chi_{\pm,k}\left(\alpha\right)=\frac{1}{\sqrt{4n}}\begin{pmatrix}
1 \\
\pm \me^{\mi\frac{ka}{2}} \\
- \me^{\mi\frac{ka}{2}} \\
\mp 1 \\
\vdots
\end{pmatrix}
\label{eqn:armchairchis}
\end{equation}
where the atoms in the 1D unit cell are labelled in order of their
angular positions around the tube.

For a finite length tube, the wavefunction is a standing wave, chosen
to satisfy the boundary conditions at the two ends of the \nt,
which places restrictions on the allowed values of $k$.  The armchair
finite tube wavefunctions take the form 
\begin{equation}
\langle i, \alpha| \Psi_{\mu ,k}\rangle =\frac{\me^{\mi\frac{ka}{2}}}{{C}}\left[\me^{-i\frac{ka}{2}}\me^{\mi kx_{i}}\chi_{\mu ,k}\left(\alpha
\right)-\me^{\mi\frac{ka}{2}}\me^{-ikx_{i}}\chi^{*}_{\mu ,k}\left(\alpha
\right)\right] \label{eq:postphd1}
\end{equation}
where $C=\sqrt{8 n N}$ is a normalisation factor, which is independent of $k$ for a long tube.

Substituting the wavefunctions and energy dispersion relation into
Eqn.~\ref{eqn:gfsumdefapp} we can convert the summation to an
integral, and using symmetry relations and the fact that the 
resultant integral is analytic at $k=0$ we find
\begin{widetext}
\begin{equation}
\left<i,\alpha \right| G \left| j,\beta \right> =\frac{1}{8\pi
n}\sum_{\mu}\int_{-2\pi}^{2\pi}d(ka)\frac{\me^{\mi ak(i-j)}\chi_{\mu ,k}(\alpha)\chi_{\mu ,k}^{*}(\beta)}{E-\epsilon_{\mu ,k}+i\delta}-\frac{\me^{\mi ak(i+j-1)}\chi_{\mu ,k}(\alpha)\chi_{\mu ,k}(\beta)}{E-\epsilon_{\mu ,k}+i\delta}
\label{eq:postphd2}
\end{equation}
\end{widetext}
The calculation simplifies further by considering only the Green's
function at the end of the \nt , i.e. $i=j=1$. (This is all that is
required in our calculations of the influence of the semi-infinite
leads in \S\ref{sec:gf}.)
The dependency on the inner product in Eqn.~\ref{eq:postphd2}
(describing the coupling between the different sites labelled by
$\alpha$) can be reduced to a $2\times 2$ matrix, since there are only
two types of atom on the ``edge'' of the armchair nanotube.  Using
contour integration in the variable $z=\me^{\pm\frac{ika}{2}}$, we
find the Green's function for an armchair \nt\cite{footnote5}
\begin{equation}
\left<\alpha \right| G \left| \beta
\right>=\frac{1}{2n t}\begin{pmatrix}
-z_{+}+z_{-} & z_{+}+z_{-} \\
z_{+}+z_{-} & -z_{+}+z_{-} \
\end{pmatrix}
\end{equation}
where 
\begin{eqnarray}
z_\pm&\equiv &\left(\frac{1\mp E/t}{2}\right)\pm i
\sqrt{1-\left(\frac{1\mp E/t}{2}\right)^{2}} 
\end{eqnarray}

\subsection{Zigzag Green's Functions}
The two metallic bands in an $(n,0)$ zigzag tube have modes, $\mu$,
which are distinguished by the subband indices $q_\mu=2n/3,4n/3$.
They are degenerate with energies
\begin{equation}
\epsilon_{\mu ,k}=\pm
2 t\sin{\left(\frac{\sqrt{3}ka}{4}\right)}
\end{equation}
for the right and left moving modes ($\pm$). For a tube of finite
length, the boundaries induce reflections but no mixing between the
bands of different $q$ since rotational symmetry is preserved. The
finite tube wavefunctions take the form
\begin{equation}
\langle i, \alpha
\ket{\Psi_{\mu ,k}}=\frac{\me^{\mi q_\mu\frac{2\pi}{n}y_{\alpha}}}{{C}}\left[\me^{\mi kx_i}\chi_{\mu ,k}\left(\alpha
\right)-\me^{-ikx_i}\chi_{\mu ,-k}\left(\alpha \right)\right]
\label{eq:postphdzz}
\end{equation}
with the same notation as in the armchair case (but note that the 1D
unit cell and associated wavefunctions $\chi$ are now those
appropriate for the zigzag tube).

 We proceed as in the armchair case by restricting our attention to
the relevant atoms on one end of the \nt . Converting the summation to
an integral and using residue calculus we find the total metallic
Green's function on the end of a semif-infinite zigzag \nt~ to be
\begin{widetext}
\begin{equation}
\left<\alpha\right|G\left|\beta\right>=\frac{2}{nt}\me^{\mi\frac{2\pi}{3a}\left(y_{\alpha}-y_{\beta}\right)}\left(1+\me^{\mi\frac{2\pi}{3a}\left(y_{\alpha}-y_{\beta}\right)}\right)
\left[\frac{E}{2t}-i\sqrt{1-\left(\frac{E}{2t}\right)^{2}}\right]
\end{equation}
\end{widetext}
where the terms of two periodicities in $y_\alpha-y_\beta$ arise
from the $4n/3$ and $2n/3$ bands.

\section{Transmission probabilities}

\label{app:transmissioncoefficients}

Here we state the transmission probabilities across a scattering
region between armchair and zigzag \nt s.  Following \S\ref{sec:gf}
the input wave is labelled by mode $\mu$ and lead $p$, and the output
by mode $\mu'$ and lead $p'$. One must distinguish between the cases
where the input and output leads are of the differing chiralities
\begin{widetext}
\begin{eqnarray}
\mbox{$p$ armchair, $p'$ armchair} \quad\quad T_{\mu,p;\mu',p'} & = & \left|
-\delta_{p,p'}\delta_{\mu,\mu'}+2\frac{\hbar}{at}v_{\rm ac}\left<\chi_{\mu}\right|
\left.\mathcal{G}\right.\left|\chi_{\mu'}\right>\right|^2\\
\mbox{$p$ zigzag, $p'$ zigzag} \quad\quad T_{\mu,p;\mu',p'}
 & = &
\left|-\delta_{p,p'}\delta_{\mu,\mu'}+2\frac{2\hbar}{\sqrt{3}at}v_{\rm zz}\left<\chi_{\mu}\right|
\left.\mathcal{G}\right.\left|\chi_{\mu'}\right>\right|^2\\
\mbox{$p$ armchair, $p'$ zigzag (or vice versa)} \quad\quad T_{\mu,p;\mu',p'} & = &
\left|\frac{2\hbar}{at}\sqrt{\frac{2}{\sqrt{3}}v_{\rm zz}v_{\rm ac}}\left<\chi_\mu\right|
\left.\mathcal{G}\right.\left|\chi_{\mu'}\right>\right|^2
\end{eqnarray}
\end{widetext}
 We use the terminology and definitions for the \nt~wavefunctions
described in Appendix~\ref{app:ntgreenfns}, and $v_{\rm ac}$ and
$v_{\rm zz}$ are the velocities for armchair and zigzag tubes.  The
inner product of the Green's function with the 1D wavefunctions
$\chi_{\mu}$ $\chi_{{\mu'}}$ are to be taken on any 1D unit cell in
the leads $p$ and $p'$ (wavevector labels are suppressed, but should
be understood to be the Fermi wavevectors of the leads).  The
Kronecker delta functions represent the modification required for
reflection back into the same lead and mode.

\section{Incoherent Scattering Summation}
\label{app:inco}

\begin{figure}
\centering
\includegraphics[scale=0.3]{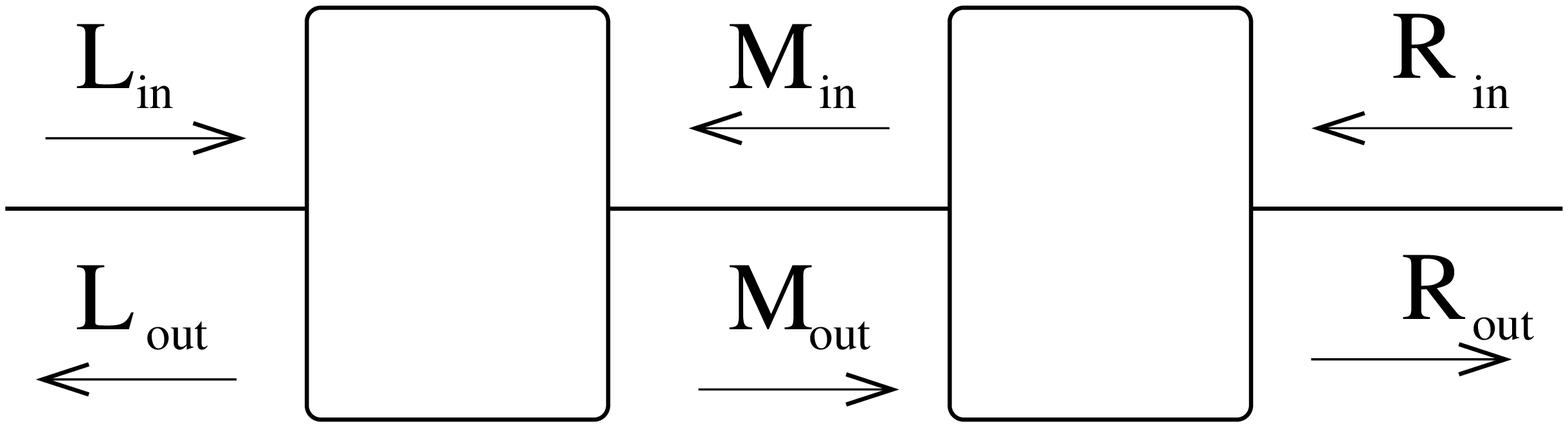}
\caption{Combining two identical scattering sections}
\label{fig:incoscatt}
\end{figure}

We illustrate the method used to combine probability $s$-matrices,
which enables us to calculate conductance through many scattering
centres incoherently, as used in \S\ref{sec:incomain}.  Consider
the setup in Fig.~\ref{fig:incoscatt}.
For two identical scattering sections we have
\begin{eqnarray}
\left( \begin{array}{c} L_{\rm out} \\ M_{\rm out} \end{array}\right)   & = &
\left( \begin{array}{cc} \mathbf{R} & \mathbf{T} \\ \mathbf{T} &
    \mathbf{R} \end{array} \right)\left( \begin{array}{c} L_{\rm in} \\
  M_{\rm in} \end{array}\right) \\
\left( \begin{array}{c} M_{\rm in} \\ R_{\rm out} \end{array}\right)   & =
& \left( \begin{array}{cc} \mathbf{R} & \mathbf{T} \\ \mathbf{T} &
    \mathbf{R} \end{array} \right)\left( \begin{array}{c} M_{\rm out} \\
  R_{\rm in} \end{array}\right)
\end{eqnarray}
where $L_{\rm out}$ is a vector representing the probability flux of
all the outgoing modes to the left of the scatterer, and $L_{\rm in}$
is a vector representing the probability flux of all the ingoing modes
to the left, similarly for $M$ and $R$. $\mathbf{R}$ and $\mathbf{T}$
are matrices representing the reflection and transmission of different
leads. We assume that the scattering sections are symmetric in
space. We can eliminate the $M$ terms to obtain the combined structure
\begin{equation}
\left( \begin{array}{c} L_{\rm out} \\ R_{\rm out} \end{array}\right)  =
\left( \begin{array}{cc} \mathbf{R}' & \mathbf{T}' \\ \mathbf{T}' &
    \mathbf{R}' \end{array} \right)\left( \begin{array}{c} L_{\rm in} \\
  R_{\rm in} \end{array}\right)
\end{equation}
where
\begin{eqnarray}
\mathbf{T}'=\mathbf{T}\left[ \mathbf{I}-\mathbf{R}^{2}
  \right]^{-1} \mathbf{T} \\
\mathbf{R}'=\mathbf{R} + \mathbf{T}\mathbf{R}\left[
  \mathbf{I}-\mathbf{R}^{2} \right]^{-1} \mathbf{T}
\end{eqnarray}
For the setup with two \nt s, it is useful to separate
tubes A and  B:
\begin{equation}
\left( \begin{array}{c} L^{\rm A}_{\rm out} \\ R^{\rm A}_{\rm out}  \\
  L^{\rm B}_{\rm out} \\ R^{\rm B}_{\rm out} \end{array}\right)  =
\left( \begin{array}{cccc}
\mathbf{R}^{\rm A} & \mathbf{T}^{\rm A} & \mathbf{I} & \mathbf{I}\\
  \mathbf{T}^{\rm A}  &
    \mathbf{R}^{\rm A}  & \mathbf{I} & \mathbf{I} \\
\mathbf{I} & \mathbf{I} & \mathbf{R}^{\rm B} & \mathbf{T}^{\rm B} \\
\mathbf{I} & \mathbf{I} & \mathbf{T}^{\rm B} & \mathbf{R}^{\rm B}
\end{array}\right)
\left( \begin{array}{c} L^{\rm A}_{\rm in} \\ R^{\rm A}_{\rm in}  \\
  L^{\rm B}_{\rm in} \\ R^{\rm B}_{\rm in} \end{array}\right)
\end{equation}
where $\mathbf{I}$ is the intertube transmission probability, which
has been set equal between all modes to simplify the problem. Taking
one mode per tube, adding $N_{\rm s}$ identical scatterers, and
expanding in a power series in terms of $I$, the intertube
transmission probability, the leading behaviour of the relevant
probabilities is
\begin{eqnarray}
T_{N_{\rm s}} & =& \frac{T}{N_{\rm s}-(N_{\rm s}-1)T} \quad \lim_{N_{\rm s}\rightarrow
  \infty}\rightarrow  \frac{T}{N_{\rm s}(1-T)} \\
I_{N_{\rm s}}& =& N_{\rm s}I
\end{eqnarray}
where $T_{N_{\rm s}}$ is the total intratube transmission probability
through $N_{\rm s}$ scatterers if $T$ is the probability of transmission
through one scatterer, and $I_{N_{\rm s}}$ is the total intertube
transmission probability through $N_{\rm s}$ scatterers if $I$ is the
probability of transmission through one scatterer.  We make use of
these results in \S\ref{sec:incomain} in the limit of weak
scattering $(1-T), I \ll 1$.


\begin{thebibliography}{10}
\bibitem{footnote1} Here we focus on momentum conservation along the tubes,
which is all that is conserved in the case of the rope geometry. We
defer a discussion of rotational symmetry in multiwall tubes to
\S\ref{sec:mwsym}.
\bibitem{footnote2} This result is in contrast to the results in
Ref.~\onlinecite{chico}, where it was suggested that the conductance
at $E=0$ grows with the circumference of the \nt.
\bibitem{footnote3} Our results on the discrete lattice sample length
at the values $L=\mu a$ with $\mu$ integer, thus there is an arbitrary
offset of $\pi/a$.
\bibitem{footnote4} $R_s$ is one half of the total reflection
coefficient. Given that there are two incoming modes, it can be
interpreted as the average reflection probability for each mode.
\bibitem{dents} Owing to uncertainties in the experimental parameters
we cannot rule out some contribution of deformations of the outer
shell (which will act to reduce the ratio $R_s/I_s$) in addition to
potential defects.
\bibitem{footnote5} Here $\alpha$ and $\beta$ are restricted to the
two atomic positions on the edge of the armchair nanotube of differing
local symmetry.
\end{thebibliography}
\end{document}